\documentclass[aps,twocolumn,showpacs,superscriptaddress,groupedaddress,amssymb,amsfonts,amsmath,letterpaper]{revtex4-2}  

\usepackage{graphicx}
\usepackage{amsmath}
\usepackage{upgreek}
\usepackage{cancel}
\usepackage{natbib}
\usepackage{color}
\usepackage{hyperref}
\definecolor{darkblue}{rgb}{0.0,0.0,0.3}
\hypersetup{colorlinks,breaklinks,
            linkcolor=darkblue,urlcolor=darkblue,
            anchorcolor=darkblue,citecolor=darkblue}

\newcommand{\mnras}{Mon.~Not.~Roy.~Astron.~Soc.}
\renewcommand{\pop}{Phys.~Plasmas}
\renewcommand{\pof}{Phys.~Fluids}
\newcommand{\apjl}{Astrophys.~J.~Lett.}

\newcommand{\pasj}{Proc.~Astron.~Soc.~Japan}

\newcommand{\pfluid}{Phys.~Fluids}
\renewcommand{\apj}{Astrophys.~J.}
\newcommand{\apjs}{Astrophys.~J.~Supp.}

\newcommand{\astrnat}{Nat.~Astron.}
\renewcommand{\jpp}{J.~Plasma Phys.}

\newcommand{\ssr}{Space Sci. Rev.}

\renewcommand{\prl}{Phys.~Rev.~Lett.}
\newcommand{\aap}{Astron.~Astro.}

\newcommand{\pr}{Phys.~Rev.}
\newcommand{\solphys}{Solar Phys.}

\newcommand{\jcomp}{J.~Comp.~Phys.}

\newcommand{\jgr}{J.~Geophys.~Res.}
\newcommand{\grl}{Geophys.~Res.~Lett.}

\newcommand{\lrsp}{Living Rev.~Solar Phys.}

\newcommand{\pnas}{Proc.~Nat.~Acad.~Sci.}

\newcommand{\dansssr}{Dok.~Akad.~Nauk.~SSSR}

\newcommand{\ropp}{Rev.~Plasma Phys.}

\newcommand{\rspa}{Proc.~R.~Soc.~Lond.}

\newcommand{\jetp}{Sov.~J.~Exp.~Theor.~Phys.} 

\newcommand{\frontiers}{Front.~Astron.~Space Sci.}

\newcommand{\D}[2]{\frac{{\rm d} #2}{{\rm d} #1}}

\newcommand\bb[1]{\mbox{\boldmath{$#1$}}}
\newcommand\grad{\bb{\nabla}}
\newcommand\bcdot{\,\bb{\cdot}\,}
\newcommand\bdbldot{\,\bb{:}\,}
\newcommand\btimes{\,\bb{\times}\,}

\newcommand{\msb}[1]{\bb{\mathsf{#1}}}
\newcommand\bs[1]{\boldsymbol{#1}}

\newcommand{\ez}{\hat{\bb{z}}}
\newcommand{\eb}{\hat{\bb{b}}}

\newcommand{\ROS}{\eb\eb\bdbldot\grad\bb{u}}

\begin{document}

\title{Kinetic Turbulence in Collisionless High-Beta Plasmas}

\author{Lev Arzamasskiy}\email{leva@ias.edu}
\affiliation{School of Natural Sciences, Institute for Advanced Study, Princeton, NJ 08540, USA}
\author{Matthew W.~Kunz}
\affiliation{Department of Astrophysical Sciences, 
Princeton University, Peyton Hall, Princeton, NJ 08544, USA}
\affiliation{Princeton Plasma Physics Laboratory, PO~Box 451, Princeton, NJ 08543, USA}
\author{Jonathan Squire}
\affiliation{Physics Department, University of Otago, 730 Cumberland St, North Dunedin, Dunedin 9016, New Zealand}
\author{Eliot Quataert}
\affiliation{Department of Astrophysical Sciences, 
Princeton University, Peyton Hall, Princeton, NJ 08544, USA}
\author{Alexander A.~Schekochihin}
\affiliation{The Rudolf Peierls Centre for Theoretical Physics, University of Oxford, Clarendon
Laboratory, Parks Road, Oxford OX1 3PU, UK}
\affiliation{Merton College, Merton Street, Oxford OX1 4JD, UK}

\date{\today}

\begin{abstract}
We present results from three-dimensional hybrid-kinetic simulations of Alfv\'enic turbulence in a high-beta, collisionless plasma. The key feature of such turbulence is the interplay between local wave--wave interactions between the fluctuations in the cascade and the non-local wave--particle interactions associated with kinetic micro-instabilities driven by anisotropy in the thermal pressure (namely, firehose, mirror, and ion-cyclotron). We present theoretical estimates for, and calculate directly from the simulations, the effective collisionality and plasma viscosity in pressure-anisotropic high-beta turbulence, demonstrating that, for strong Alfv\'enic turbulence, the effective parallel-viscous scale is comparable to the driving scale of the cascade. Below this scale, the kinetic-energy spectrum indicates an Alfv\'enic cascade with a slope steeper than $-5/3$ due to the anisotropic viscous stress. The magnetic-energy spectrum is shallower than $-5/3$ near the ion-Larmor scale due to fluctuations produced by the firehose instability. Most of the cascade energy (${\approx}80$--90\%) is dissipated as ion heating through a combination of Landau damping and anisotropic viscous heating. Our results have implications for models of particle heating in low-luminosity accretion onto supermassive black holes, the effective viscosity of the intracluster medium, and the interpretation of near-Earth solar-wind observations.

\end{abstract}

\maketitle

\section{Introduction}
\label{sec:introduction}

Many space and astrophysical plasmas are so hot and dilute that the mean free path between particle--particle binary interactions is comparable to (or even larger than) the characteristic scales of the system. Examples of such weakly collisional plasmas include the intracluster medium (ICM) of galaxy clusters (characteristic scale $L\sim 100~{\rm kpc}$, Coulomb-collisional mean free path $\lambda_{\rm mfp}\sim 1{-}10~{\rm kpc}$~\citep{Zhuravleva2015}), low-luminosity accretion flows onto super-massive black holes (e.g., for the accretion flow around Sgr~A$^\star$ at the Bondi radius, $\lambda_{\rm mfp} \sim L \sim 0.1~{\rm pc}$~\citep{Baganoff2003}), and the near-Earth solar wind ($\lambda_{\rm mfp} \sim L \sim 1~{\rm au}$~\citep{MarschGoldstein1983}). All of these plasmas also have particle Larmor radii $\rho$ many orders of magnitude smaller than macroscopic scales (e.g., $\rho/L \lesssim 10^{-14}$ for the ICM, ${\lesssim}10^{-10}$ for Sgr A$^{\star}$, and ${\lesssim}10^{-6}$ for the solar wind). Despite this strong magnetization, the magnetic fields in these systems are typically energetically sub-dominant, with the ratio of thermal pressure $p$ and magnetic pressure $B^2/8\pi$, $\beta \equiv 8\pi p/B^2 \gtrsim 1$.

A particularly interesting question in such high-$\beta$, low-collisionality plasmas is how the kinetic physics, which acts on extremely small (and often unobservable) scales, influences the global evolution of the system and impacts the interpretation of current and future observations. The deviations from local thermodynamic equilibrium allowed by the low collisionality of these plasmas can have a dramatic effect on the transport of energy and momentum and the evolution of cosmic magnetic fields. For example, the viscous stress caused by velocity-space anisotropy in the particle distribution function can provide an order-unity contribution to the mass-accretion rate, enhancing or reducing it depending on the shape of the particle distribution~\citep{Sharma2006,Kunz2016,Foucart2017,Kempski2019}. In addition, the electrons may have a different temperature than the poorly radiating ions~\citep{NarayanYi1994,NarayanYi1995}, thereby complicating the interpretation of interferometric images of black-hole accretion flows (such as those taken by the Event Horizon Telescope~\cite{EHT1,EHT_SgrA5}). As a result, in order to understand high-$\beta$ astrophysical systems, one must discover how the energy injected in these systems by large-scale processes gets transferred to the electron and ion distribution functions, and how these distributions are shaped by field-particle interactions and various kinetic instabilities.  

In this paper, we explore energy transfer and dissipation in collisionless high-$\beta$ turbulence. This problem is of fundamental importance for the aforementioned astrophysical systems because all of them are observed or thought to host a broadband cascade of turbulent fluctuations. Most theories of magnetohydrodynamic (MHD) turbulence assume that the local non-linear turbulent interactions are the main process by which energy is transferred from large scales to kinetic scales (see Ref.~\cite{Schekochihin2020} for a recent review). In collisionless systems, additional energy-transfer channels are available, not only via phase mixing to small scales in velocity space~\citep{Schekochihin2009} but also through the action of a number of kinetic micro-instabilities at ion (firehose~\citep{Rosenbluth1956, Parker1958b, Chandrasekhar1958,VedenovSagdeev1958}, mirror~\citep{ShapiroShevchenko1964,Barnes1966,SouthwoodKivelson1993}, and Alfv\'en ion-cyclotron~\citep[AIC,][]{SagdeevShafranov1960}) and electron (firehose, mirror, and whistler~\citep{SagdeevShafranov1960}) scales. These instabilities feed off the free energy associated with field-anisotropic deviations from local thermodynamic equilibrium (e.g., pressure anisotropies). In the absence of collisions (including any effective collisionality due to wave--particle interactions), such deviations are expected to be largest within the injection range of the cascade, where the amplitude of the turbulent motions is largest \footnote{In making this statement, we implicitly assume conservation of adiabatic invariants. We discuss this in more detail in \S\ref{sec:theory}.}. The implication is a non-local transfer of energy from large ``fluid'' scales to astrophysically microscopic kinetic scales, instead of a scale-by-scale cascade that is customarily believed to occur in more mundane systems~\cite{GS1995}. This is not necessarily the case in weakly collisional plasmas, in which the (effective) collisionality is large enough to efficiently isotropize the distribution at the injection scale but not throughout the inertial range. In such plasmas, the pressure anisotropy is driven by the motions between the injection scale and the effective viscous scale, the latter of which being mediated by the wave--particle interactions from micro-instabilities. As we show in \S\ref{sec:theory}, this effective viscous scale can be comparable to the injection scale of the cascade, making the interactions between micro-instabilities and the cascade crucial for the dynamics even at macroscopic scales.

The effect of kinetic micro-instabilities on the plasma has been studied with pressure anisotropies driven either externally (with large-scale shear~\citep{Kunz2014,Riquelme2015} or expansion~\citep{Hellinger2008,Sironi2015}) or by individual nonlinear waves (Alfv\'en~\citep{Squire2017a}, ion-acoustic~\citep{Kunz2020}, magnetosonic~\citep{Majeski2023}). Here we explore, for the first time, and using six-dimensional hybrid-kinetic numerical simulations, the interaction between the local cascade of strong turbulence and the non-local excitation of microscale kinetic instabilities that are self-consistently produced by the fluctuations in the cascade themselves.

The article is organized as follows. We begin in~\S\ref{sec:theory} with a summary of analytical and numerical results on the properties of waves and turbulence in pressure-anisotropic, high-$\beta$ plasmas. We use these results to obtain analytical estimates for the effective collisionality and effective viscosity in such plasmas. We then test these results with self-consistent numerical simulations in~\S\ref{sec:numerics}, which allow us to measure the effective collisionality and viscosity and to determine the dominant energy dissipation mechanisms in collisionless high-$\beta$ turbulence. In~\S\ref{sec:discussion} our results and their range of validity are discussed and put in the context of observations of turbulence in the ICM and in the solar wind. We close in~\S\ref{sec:summary} with a summary of our results.

\section{Theoretical expectations}
\label{sec:theory}

\begin{figure*}
    \centering
    \includegraphics[width=\textwidth]{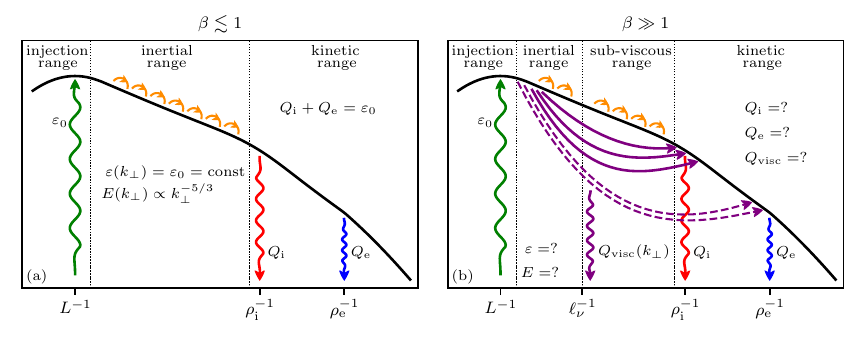}
    \caption{Qualitative picture of how the cascade proceeds in low-$\beta$ (a) and high-$\beta$ (b) kinetic turbulence. For $\beta \lesssim 1$, the energy flux, injected at some large scale $L$, remains constant in the inertial range, and is eventually dissipated by ions and electrons at the corresponding kinetic scales. In contrast, high-$\beta$ plasma allows for the non-local energy transfer by kinetic micro-instabilities. The effective viscosity of such a plasma can convert bulk kinetic energy into thermal energy. The goal of this paper is to examine the effects of this physics on the turbulent cascade and on the distribution of energy between species.}
    \label{fig:schematic}
\end{figure*}

Modern theories of strong plasma turbulence in magnetized plasmas (starting with~\citet{GS1995}) stipulate a dynamical balance between the linear physics of plasma fluctuations and their nonlinear interactions. In Alfv\'enic turbulence, the typical linear timescale for a given eddy is determined by the time it takes for an Alfv\'en wave to cross that eddy. For an eddy whose extent along the local mean magnetic field $\bb{B}_0$ is $\ell_\parallel$, this timescale is $\tau_{\rm lin} \sim \ell_\parallel/v_{\rm A0}$, where $v_{\rm A0}$ is the associated Alfv\'{e}n speed. The nonlinear timescale, on which the mutual shearing and advection of the fluctuations decorrelates the eddy, is estimated as $\tau_{\rm nl}\sim \ell_\perp/\delta u_\perp \sim \ell_\perp/(v_{\rm A0} \,\delta B_\perp/B_0)$, where $\ell_\perp$ is the size of the fluctuation across (``perpendicular'' to) $\bb{B}_0$, and $\delta u_\perp$ and $\delta B_\perp$ are the amplitudes of the fluid-velocity and magnetic-field perturbations. The relationship between linear and non-linear timescales determines how many non-linear interactions are required to decorrelate the eddies. If $\tau_{\rm nl} \ll \tau_{\rm lin}$, different parts of an eddy decorrelate before they can be in causal contact via Alfv\'en-wave propagation, which leads to a decrease in $\ell_\parallel$ and thus $\tau_{\rm lin}$~\citep{Boldyrev2005}. On the other hand, if $\tau_{\rm nl} \gg \tau_{\rm lin}$, the turbulence is considered to be ``weak'', and it evolves in a way such that these timescales become comparable to one another, $\tau_{\rm nl} \sim \tau_{\rm lin}$ at small scales \citep{Meyrand2016}. This causes the system ultimately to settle into a scale-by-scale ``critical balance'' between the linear and nonlinear timescales~\citep{GS1995,Mallet2015}. 
The result is a \citet{Kolmogorov1941} spectrum in the direction perpendicular to the magnetic field, $k^{-5/3}_\perp$, where $k_\perp\sim 1/\ell_\perp$ is the perpendicular wavenumber,  shown schematically in Figure~\ref{fig:schematic}a. The associated scale-dependent anisotropy, $\ell_\parallel \propto \ell_\perp^{2/3}$, was confirmed numerically by Refs.~\cite{Cho2000} and \cite{MaronGoldreich2001} and subsequently measured in the solar wind using spacecraft data by Ref.~\cite{Horbury2008} (and many others; see~\cite{BrunoCarbone2013} and~\cite{Chen2016} for reviews).

In collisionless plasmas, non-linear wave--wave interactions are accompanied by additional linear and non-linear physics related to wave--particle interactions. For example, changes in magnetic-field strength caused by turbulent fluctuations on scales much larger than the kinetic scales lead to changes in perpendicular pressure through the conservation of particles' magnetic moments~\citep{Chew1956}. The pressure then becomes {\em anisotropic} with respect to the local magnetic-field direction, with the field-perpendicular component of the pressure, $p_\perp$, differing from the field-parallel component of the pressure, $p_\parallel$. This pressure anisotropy effectively modifies the Alfv\'en speed by contributing a field-aligned viscous stress~\cite{Parker1957,Parker1958a,Braginskii1965}, thereby changing the characteristic linear timescale that features in the ``critical balance''~\cite{Kunz2018,Bott2021}. To see that, consider the equation for the evolution of the fluid velocity $\bb{u}$ in the presence of pressure anisotropy,
\begin{equation}\label{eqn:euler}
    \varrho \D{t}{\bb{u}} = -\grad\bcdot\biggl( \msb{P} + \frac{B^2}{8\pi} \msb{I} - \frac{B^2}{4\pi}\eb\eb \biggr),
\end{equation}
where ${\rm d}/{\rm d}t \equiv \partial/\partial t + \bb{u}\bcdot\grad$ is the comoving time derivative, $\varrho$ is the mass density, $\msb{P}$ is the pressure tensor, $\msb{I}$ is the unit dyadic, and $\eb\equiv\bb{B}/B$ is the unit vector in the direction of the magnetic field $\bb{B}$. For a magnetized plasma in which the characteristic timescales are much longer than ion-Larmor period $2\pi/\Omega_{\rm i}$, the pressure tensor is predominantly diagonal in a coordinate frame defined by the field direction~\cite{Chew1956}:
\begin{equation}
    \msb{P} = p_\perp \bigl(\msb{I} - \eb\eb\bigr) + p_\parallel \eb\eb = p_\perp \msb{I} - \Delta p \, \eb\eb ,
\end{equation}
the latter equality defining the pressure anisotropy $\Delta p \equiv p_\perp - p_\parallel$. Equation \eqref{eqn:euler} may then be rewritten as
\begin{equation}\label{eqn:euler2}
    \varrho \D{t}{\bb{u}} = -\grad \biggl(\frac{B^2}{8 \pi} + p_\perp\biggr) - \grad\bcdot \biggl[ \biggl(\frac{B^2}{4 \pi} + \Delta p\biggr) \eb \eb \biggr].
\end{equation}
The final term in equation~\eqref{eqn:euler2} highlights the role of the pressure anisotropy in modifying the magnetic tension force compared to MHD, {\em viz.},
\begin{equation}
\frac{\bb{B}\bcdot\grad\bb{B}}{4\pi} = \grad\bcdot\left[\frac{B^2}{4\pi}\eb\eb\right] \rightarrow \grad\bcdot\left[\left(\frac{B^2}{4\pi}+\Delta p\right)\eb\eb\right].
\end{equation}
As a result of this modification, the effective Alfv\'{e}n speed in the plasma,
\begin{equation}\label{eqn:vAeff}
    v_{\rm A,eff} \equiv v_{\rm A} \biggl( 1 + \frac{\beta}{2} \Delta \biggr)^{1/2},
\end{equation}
where $\Delta \equiv \Delta p/p$, $\beta\equiv 8\pi p/B^2$ and $p = (2p_\perp + p_\parallel)/3$, may depart significantly from $v_{\rm A}$ when $|\Delta| \sim 1/\beta$. In particular, as the firehose instability threshold $\Delta = -2/\beta$, below which $v_{\rm A,eff}$ becomes imaginary, is approached, it becomes energetically ``cheaper'' for the fluid motions to bend the magnetic-field lines~\cite{Kunz2015}. Thus, with $\beta\gg{1}$, even small departures from pressure isotropy can influence the plasma dynamics in a dramatic way.

What makes this influence particularly complicated in a turbulent environment is the associated spatio-temporal inhomogeneity of the pressure anisotropy. Pressure anisotropy is generated by the (approximate) conservation of each particle's adiabatic invariants, $\mu\equiv m w^2_\perp/2B$ and 
$\mathcal{J} \equiv \oint m \bb{w}_\parallel\bcdot {\rm d}\bb{x}$, where $n$ is plasma number density, $m$ is the particle mass, and $\bb{w}_{\perp,\parallel}$ are the velocities of the peculiar (``thermal'') motions of the particle perpendicular and parallel to the local magnetic field. As the magnetic-field strength fluctuates, the perpendicular and parallel energies of the particles therefore fluctuate as well, resulting in a pressure anisotropy that varies both in space and time. To describe this evolution, if only heuristically, we use the \citet{Chew1956} equations with $|\Delta|\lesssim 1/\beta\ll{1}$ to write, to the lowest order in $\Delta$,
\begin{equation}
    \D{t}{\Delta} \approx 3\ROS - 3\nu \Delta.\label{eqn:cgl}
\end{equation}
Here we have assumed incompressible motions, included the isotropizing effect of collisions not present in the original equations, and neglected contributions from heat fluxes. The first term on the right-hand side of equation~\eqref{eqn:cgl} captures the adiabatic production of pressure anisotropy caused by changes in magnetic-field strength as measured in the fluid frame. Indeed, adopting an ideal Ohm's law and assuming incompressibility, Faraday's law of induction provides ${\rm d}\ln B/{\rm d}t = \ROS$. In this case, pressure anisotropy is driven by field-parallel gradients of field-parallel flows, i.e., the ``parallel rate of strain.'' The second term in equation~\eqref{eqn:cgl} represents the relaxation of the pressure anisotropy by collisions, a process that isotropizes the distribution function on a characteristic timescale $\nu^{-1}$. In collisionless plasmas, such isotropization may be provided by the field--particle interactions, and may depend on the local distribution function and magnetic-field strength.

Adapting the Goldreich--Sridhar theory to account for the effective Alfv\'en speed (\ref{eqn:vAeff}) only works as long as the pressure anisotropy is small compared to $1/\beta$. For example, an attempt to construct such a theory in the gyrokinetic limit was made in Refs.~\cite{Kunz2015,Kunz2018}; in this theory, background pressure anisotropy modifies the fluctuations and their nonlinear interactions, while the pressure anisotropy associated with the fluctuations is too small to feed back nonlinearly on the fluctuations themselves. If instead the pressure anisotropy (either background or fluctuation-driven) exceeds any of the thresholds of the various kinetic micro-instabilities (at high plasma $\beta$, mostly firehose and mirror), it can cause energy to be transferred non-locally (see Figure~\ref{fig:schematic}b for a schematic picture of the cascade). The behavior of turbulence in such a situation is the topic of this paper. In the remainder of this section, we provide analytical estimates for the effective collisionality and viscosity of a high-$\beta$, kinetically unstable plasma supporting a turbulent cascade of electromagnetic fluctuations.

We begin by supposing that energy is injected in the form of bulk motions at an outer scale $L$ and initiates a turbulent cascade of Alfv\'enically polarized fluctuations with amplitudes $\delta u_\perp$ and $\delta B_\perp\sim (\delta u_\perp/v_{\rm A}) B_0$, where $v_{\rm A}$ is the Alfv\'en speed associated with the mean magnetic field $B_0$. We further assume that these fluctuations satisfy $\delta B_\perp / B_0 \gtrsim 1/\sqrt{\beta}$ above some scale, so that the pressure anisotropy generated adiabatically by the fluctuating magnetic-field strength is large enough to trigger firehose and/or mirror instabilities~\citep{Squire2016,Squire2017a}. In this case, the instabilities grow to wrinkle the magnetic field sharply on ion-Larmor scales, ultimately leading to pitch-angle scattering of ions at an effective collision frequency $\nu_{\rm eff}$ that is large enough to limit the fluctuating pressure anisotropy to marginally (un)stable values, viz., $|\Delta p|/p \sim |\ROS|/\nu_{\rm eff} \sim 1/\beta$~\citep{Kunz2014,Melville2016}. With $|\ROS| \sim \omega_{\rm A} (\delta B_\perp/B_0)^2$ for Alfv\'enic fluctuations that have a linear frequency $\omega_{\rm A}$, the effective collision frequency $\nu_{\rm eff}$ then satisfies
\begin{equation}\label{eqn:nueff1}
    \nu_{\rm eff} \sim \beta \,\omega_{\rm A} \frac{\delta B^2_\perp}{B^2_0} .
\end{equation}
For a critically balanced cascade, $\omega_{\rm A}$ is always comparable to the inverse of the turnover time at each scale. Namely, for an Alfv\'enic fluctuation with parallel extent $\ell_\parallel$ and perpendicular extent $\ell_\perp$, we have  $\omega_{\rm A} \sim v_{\rm A}/\ell_\parallel \sim \delta u_\perp/\ell_\perp \sim (v_{\rm A}/\ell_\perp) (\delta B_\perp/B_0)$. Equation~\eqref{eqn:nueff1} then becomes
\begin{equation}\label{eqn:nueff2}
    \nu_{\rm eff} \sim \beta \, \frac{v_{\rm A}}{\ell_\perp} \frac{\delta B^3_\perp}{B^3_0} .
\end{equation}
If the cascade is approximately conservative \footnote{If kinetic microinstabilities are active, the cascade can only be approximately conservative. In our derivation, we assume the energy leak due to instabilities to be small.}, then we furthermore have $\delta B^3_\perp \ell^{-1}_\perp \sim {\rm const}$, so that Equation~\eqref{eqn:nueff2} implies an effective collision frequency independent of $\ell_\perp$ and large enough to regulate the pressure anisotropies generated at all scales in the cascade. Evaluating Equation~\eqref{eqn:nueff2} at the outer scale $L$, we find that 
\begin{equation}\label{eqn:nueff}
    \nu_{\rm eff} \sim \beta \, \frac{v_{\rm A}}{L} \frac{\delta B^3_{\perp,L}}{B^3_0} \sim \beta \,\frac{v_{\rm A}}{L} {\rm M}_{\rm A}^3 ,
\end{equation}
where ${\rm M}_{\rm A}\equiv \delta u_L/v_{\rm A}$ is the Alfv\'enic Mach number of the outer-scale motions. 

The collisionality given by \eqref{eqn:nueff} implies an effective (field-parallel) Reynolds number
\begin{equation}
   {\rm Re}_{\parallel\rm eff} \equiv \frac{\delta u_L L}{v_{\rm thi}^2/\nu_{\rm eff}} \sim {\rm M}_{\rm A}^4.\label{eq:Re_eff}
\end{equation}
From this effective Reynolds number, one can define an effective viscous scale $\ell_{\nu} \equiv L\,{\rm Re}_{\parallel\rm eff}^{-3/4}$, which is appropriate if the dissipation rate due to the effective viscosity is proportional to $\nabla^2 \delta u^2$ \footnote{A similar derivation of the effective viscous scale in a plasma dynamo whose viscosity is controlled by kinetic microinstabilities can be found in Ref.~\cite{StOnge2020}.}. Equation~\eqref{eq:Re_eff} then implies
\begin{equation}\label{eqn:lvisc}
\ell_\nu/L \sim {\rm M}_{\rm A}^{-3}.
\end{equation}
The viscous scale given by Equation~\eqref{eqn:lvisc} is equivalent to the Alfv\'{e}n scale on which the turbulent velocity is approximately Alfv\'{e}nic. As a result, {\it for strong Alfv\'enic turbulence, the collisionless viscous scale due to kinetic micro-instabilities is comparable to the outer scale of the turbulence}. This conclusion does not depend on the plasma $\beta$ or on the exact instability that regulates the pressure anisotropy (so long as its threshold is proportional to $1/\beta$ and it is saturated via an enhancement of the plasma's collisionality).

One of the caveats of the above derivation is that we assume asymptotic scalings for the cascade, which are valid only for scales much smaller than the driving scale and much larger than the dissipation scale (i.e., in the inertial range). In this case, we interpret any implied value of $\ell_\nu > L$ as indicating $\ell_\nu \sim L$. Indeed, our results suggest that the dissipation scale (${\sim}\ell_\nu$) is comparable to the forcing scale (${\sim}L$) when ${\rm M}_{\rm A}\sim 1$. An alternative derivation of the viscous scale could be obtained by assuming $k_\parallel \delta u_\parallel \sim k_\perp \delta u_\perp$, which is equivalent to incompressibility for ${\rm M}_{\rm A} \sim 1$ fluctuations. With this assumption, $|\eb\eb\,\bb{:}\,\grad\bb{u}|_k\sim k_\perp \delta u_\perp$ increases with $k_\perp$ for a Goldreich--Sridhar-like cascade until it reaches its maximum value of $(\ell_\nu/L)^{2/3} {\rm M}_{\rm A} v_{\rm A}/L$ at $k_\perp \ell_\nu \sim 1$. This different scaling of the parallel rate of strain leads to the same value of the effective Reynolds number, ${\rm Re}_{\parallel \rm eff}\sim {\rm M}_{\rm A}^4$, implying that our results hold for non-asymptotic fluctuations. Note that our assumptions about the cascade are expected to hold only for ${\rm M}_{\rm A} \lesssim 1$. For ${\rm M}_{\rm A} \gg 1$, dynamo is expected to increase the magnetic-field strength until the Mach number decreases sufficiently~\citep{StOnge2018,StOnge2020}. In the opposite limit of ${\rm M}_{\rm A} \ll 1$, it is possible that the cascade is in the weak-turbulence regime, which has $\delta u_k \propto k_\perp^{-1/2}$ and $k_\parallel \propto {\rm const}(k_\perp)$~\citep{NgBhattacharjee1997,GoldreichSridhar1997}. Assuming such a cascade with $L_\parallel \sim L$, gives ${\rm Re}_{\parallel,{\rm eff}} \sim {\rm M}_{\rm A}^3$ and $\ell_\nu/L \sim {\rm M}_{\rm A}^{-9/4}$.

Yet another caveat to these scaling arguments is that the pressure-anisotropy stress can back-react on the motions to reduce $\hat{\bb{b}}\hat{\bb{b}}\,\bb{:}\,\grad\bb{u}$ below the simple Alfv\'enic estimate used in equation~\eqref{eqn:nueff1} (an effect termed magneto-immutability by Ref.~\citep{Squire2019}). This would reduce the drive of pressure anisotropy, and thus $\nu_{\rm eff}$ would in turn decrease compared to the above estimates. However, this effect seems unlikely to be important if $\ell_\nu$ is comparable to the scale of an external forcing, because then the dominant contribution to $\Delta p$ will be from the forcing motions rather than from somewhere in the inertial range.

To determine the importance of the effective collisionality in astrophysical systems, we use equation~\eqref{eqn:nueff} to compute the  effective ion mean free path,
\begin{equation}
    \lambda_{\parallel{\rm mfp,eff}} \sim \frac{v_{\rm thi}}{\nu_{\rm eff}} \sim {\rm M}_{\rm A}^{-2} \frac{v_{\rm thi}L}{\beta\delta u_L} \sim L \,\frac{{\rm M}_{\rm A}^{-3}}{\sqrt{\beta}}.
\end{equation}
In weakly collisional high-$\beta$ systems like the ICM, this effective mean free path could be smaller than the mean free path due to Coulomb collisions ($\lambda_{\rm Coulomb}$) even if the latter is smaller than the system size. For example, using physical parameters relevant to the Coma cluster of galaxies \citep{Kim1990,Bonafede2010,Zhuravleva2019},
\begin{align}\label{eqn:mfpeff}
    &\frac{\lambda_{\parallel{\rm mfp,eff}}}{\lambda_{\rm Coulomb}} \approx \frac{L\, {\rm M}_{\rm A}^{-3}/\sqrt{\beta}}{\frac{3\sqrt{2}}{4\sqrt{\pi}} \frac{T_{\rm i}^2}{n_{\rm i}\Lambda_{\rm i} e^4}} \\
    &\approx 0.05 \, \biggl(\frac{\delta u_L}{200~{\rm km~s}^{-1}}\biggr)^{-3} \biggl(\frac{L}{100~{\rm kpc}}\biggr) \biggl(\frac{B}{2~\mu{\rm G}}\biggr)^4   \nonumber \\
    & \quad\quad\quad\quad\quad\quad\quad\quad\quad \times\biggl(\frac{n_{\rm e}}{10^{-3}~{\rm cm}^{-3}}\biggr)^{-1}\biggl(\frac{T_{\rm e}}{10^8~{\rm K}}\biggr)^{-5/2}, \nonumber
\end{align}
where $\Lambda_{\rm i}$ is ion Coulomb logarithm, and the temperatures $T_{\rm i,e}$ and densities $n_{\rm i,e}$ of ions and electrons are assumed to be equal. This simple estimate predicts more than an order-of-magnitude suppression in the effective viscosity of the ICM, consistent with the observationally based conclusion by Refs.~\cite{Zhuravleva2019,Li2020}. Note that the scenario sketched out above for the effective mean free path is very sensitive to the magnetic-field strength (${\propto}B^{4}$) and can change easily by an order of magnitude given current measurement uncertainties.

\section{Numerical experiments}
\label{sec:numerics}

\subsection{Method of solution}
\label{sec:method}

Astrophysical high-$\beta$ plasmas typically have extremely large scale separations that are not computationally feasible to capture in numerical simulations. Because of our focus on the interplay between inertial-range turbulence and ion-scale kinetic instabilities, we adopt a hybrid-kinetic model, in which the ions are treated as kinetic while the electrons are assumed to be fluid-like. Such an approximation ignores the potential effects of electron-scale micro-instabilities on the cascade. We argue in~\S\ref{sec:discussion} that these instabilities are likely to be less important than ion-scale ones (particularly in the weakly collisional ICM, in which the electron pressure anisotropy is expected to be too small to produce instabilities), but this should be verified with fully kinetic simulations in the future. 

\subsubsection{Model equations}
\label{sec:equations}

The hybrid-kinetic approach that we employ assumes the plasma to be non-relativistic with all relevant scales much larger than the Debye length. Plasma on these scales is quasi-neutral, so $n_{\rm e} = Z_{\rm i} n_{\rm i}$, where $Z_{\rm i} \equiv q_{\rm i}/e$. The displacement current in Amp\`ere's law is negligible, hence 
\begin{equation}
    n_{\rm e} \bb{u}_{\rm e} = Z_{\rm i} n_{\rm i} \bb{u}_{\rm i} - \frac{c}{4\pi e} \grad\btimes\bb{B}.
\end{equation}
The electric field in hybrid kinetics is obtained from the equation of motion for the electron fluid,
\begin{equation}
    m_{\rm e} n_{\rm e} \frac{{\rm d} \bb{u}_{\rm e}}{{\rm d} t} = - \grad \bcdot \msb{P}_{\rm e} - e n_{\rm e} \left(\bb{E} + \frac{\bb{u}_{\rm e} \btimes \bb{B}}{c} \right),
\end{equation}
after neglecting electron inertia (the left-hand side) and specifying the form of the electron pressure tensor $\msb{P}_{\rm e}$. For simplicity, we assume the electrons to be isothermal and isotropic, $\msb{P}_{\rm e} = n_{\rm e} T_{\rm e} \msb{I}$ with $T_{\rm e} = T_{\rm i0} = {\rm const}$. As a result, the electric field in our model is given by
\begin{equation}
    \bb{E} = -\frac{\bb{u}_{\rm i}\btimes \bb{B}}{c} + \frac{(\grad\btimes\bb{B})\btimes \bb{B}}{4\pi e n_{\rm i} Z_{\rm i}} - \frac{T_{\rm e}\grad n_{\rm i}}{e n_{\rm i}},\label{eq:hybrid_E}
\end{equation}
where the first term is the (MHD) motional electric field, the second is associated with the Hall effect, and the third represents the thermoelectric effect. The magnetic field evolves according to the induction equation,
\begin{equation}
    \frac{1}{c}\frac{\partial \bb{B}}{\partial t} = - \grad\btimes\bb{E},\label{eq:induction}
\end{equation}
and satisfies $\grad\bcdot\bb{B}=0$. Note that the final term in \eqref{eq:hybrid_E} does not contribute to Faraday's law \eqref{eq:induction}, as it may be written as a full derivative ${\propto}\ln n_{\rm i}$ and is thus electrostatic.

The ion distribution function $f(t,\bb{x},\bb{v})$ evolves according to the collisionless Vlasov equation
\begin{align}
   \frac{\partial f}{\partial t} &+ \bb{v}\bcdot\frac{\partial f}{\partial \bb{x}} \nonumber\\*
   \mbox{} &+ \left[\frac{Z_{\rm i} e}{m_{\rm i}} \left(\bb{E} + \frac{\bb{v}\btimes\bb{B}}{c}\right) + \frac{\bb{F}}{m_{\rm i}}\right]  \bcdot \frac{\partial f}{\partial \bb{v}} = 0,\label{eq:vlasov}
\end{align}
where $\bb{F}$ is an external force that we use to drive turbulence at the largest scales of the simulation box (specified in \S\ref{sec:parameters}). The number density $n_{\rm i} (t,\bb{x})$ and flow velocity $\bb{u}_{\rm i}(t,\bb{x})$ of ions are then obtained by taking the zeroth and first moments of $f$. 

\subsubsection{\textsc{Pegasus\texttt{++}}}
\label{sec:ppp}

Equations (\ref{eq:hybrid_E})--(\ref{eq:vlasov}) are solved using a new hybrid-kinetic code, \textsc{Pegasus}\texttt{++} (Arzamasskiy et al., {\it in prep.}), which is based on the algorithms of its predecessor \textsc{Pegasus} \citep{Pegasus} and on the infrastructure of the popular magnetohydrodynamic code \textsc{Athena}\texttt{++} \citep{Athena++}. \textsc{Pegasus}\texttt{++} is much better optimized to take advantage of modern supercomputing architectures than \textsc{Pegasus}, thereby making the simulations reported in this paper possible. 

\textsc{Pegasus}\texttt{++} solves equation~(\ref{eq:vlasov}) using a particle-in-cell approach, in which the distribution function is represented with a finite number of macro-particles. These macro-particles' positions and velocities evolve along the characteristics of equation~(\ref{eq:vlasov}), with electric and magnetic fields interpolated from the computational grid to the particle positions using a second-order (triangular) shape function. The latter ensures that, in the limit of infinite resolution, the moments of the ion distribution function and their derivatives are continuous in space. \textsc{Pegasus}\texttt{++} solves equations~\eqref{eq:induction} and \eqref{eq:vlasov} with the electric field given by equation~\eqref{eq:hybrid_E} using a predictor-predictor-corrector method that is second-order accurate in both time and space. The code employs a staggered grid to preserve $\grad\bcdot\bb{B}=0$.

\subsubsection{Simulation parameters}\label{sec:parameters}

The ion macro-particles in our simulations are initialized to have a Maxwell--Boltzmann distribution with spatially uniform density $n_{\rm i} (t=0,\bb{x}) = n_{\rm i0}$ and temperature $T_{\rm i0}$. The electron temperature is constant, $T_{\rm e} = T_{\rm i0}$, and $Z_{\rm i} = 1$. The initial magnetic (``guide'') field is uniform $\bb{B} (t=0,\bb{x}) = B_0 \ez$.

\begin{figure*}
    \centering
    \includegraphics[width=\textwidth]{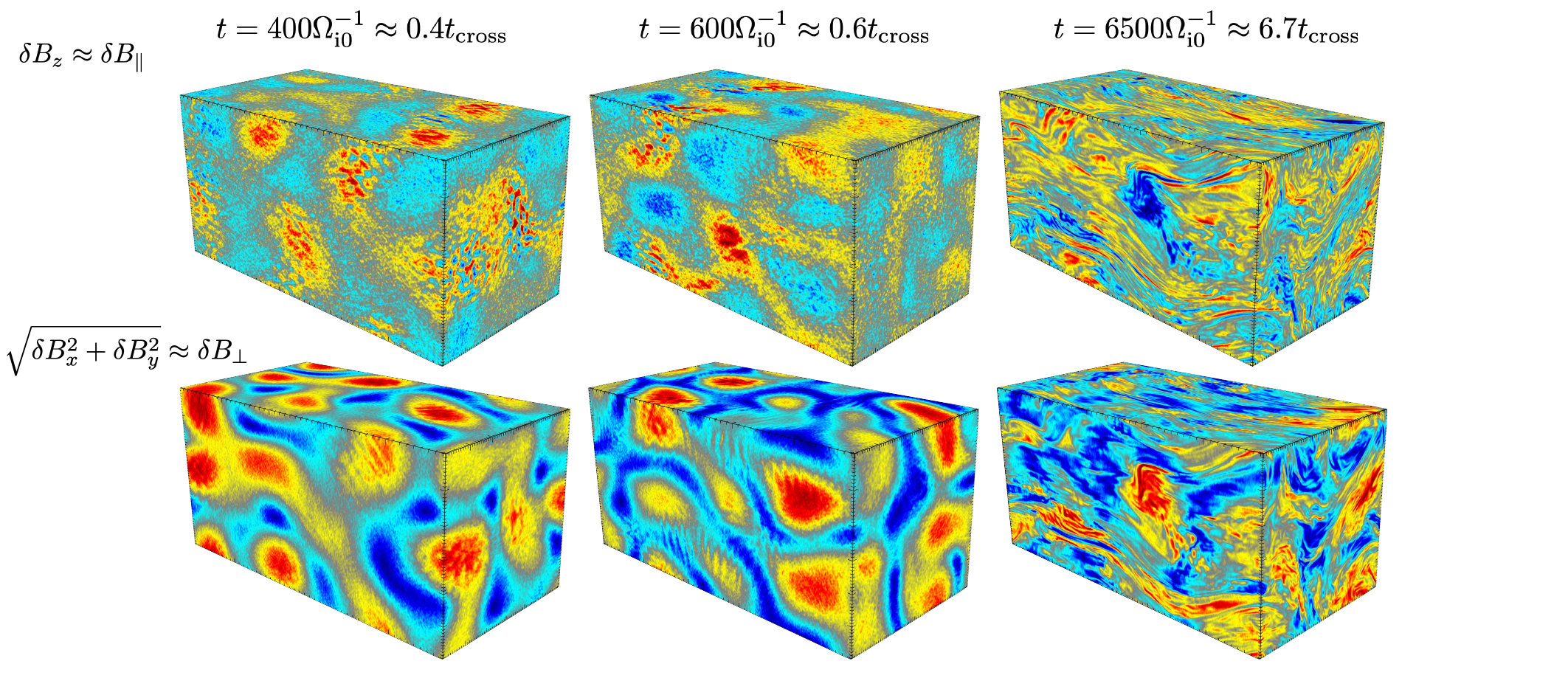}
    \caption{Time-evolution of the magnetic field strength along the guide field ($\delta B_z\approx \delta B_\parallel$, upper row), and perpendicular to the guide field ($\sqrt{\delta B_x^2 + \delta B_y^2}\approx \delta B_\perp$, lower row). We show three snapshots: $t \approx 0.4 t_{\rm cross}$, when we see the first mirror fluctuations (see also Figure~\ref{fig:slice_mirror} in Appendix~\ref{sec:appendix_instabilities} and further discussion there); $t\approx 0.6 t_{\rm cross}$, when we detect AIC fluctuations (see also Figure~\ref{fig:slice_IC}); and in the quasi-steady state ($t\approx 6.7 t_{\rm cross}$), when we can instead identify firehose fluctuations (see also Figure~\ref{fig:slice_steady_state}).}
    \label{fig:3D}
\end{figure*}

Bulk flows in the ion species are driven on large scales by an external force $\bb{F}$, which is oriented perpendicular to the guide field ($\bb{F}\perp\ez$) and constructed to be solenoidal ($\grad\bcdot\bb{F}=0$). This results in almost incompressible fluctutations with typical density variation of ${\sim}5\%$. The force is correlated in time using an Ornstein--Uhlenbeck process with a correlation time equal to the Alfv\'en crossing time, $t_{\rm corr} = (k_{\parallel}^{\rm f} v_{\rm A0})^{-1}$, associated with the smallest parallel wavenumber of the forcing $k_{\parallel}^{\rm f}$. This forcing results in fluctuations that are primarily Alfv\'enically polarized, a feature that makes our simulations relevant to many space and astrophysical plasmas. For example, observations of turbulence in the solar wind find that most of the power is in fluctuations that are Alfv\'enically polarized~\citep{BelcherDavis1971,TuMarsch1995,BrunoCarbone2005}. Outside of the solar wind, {\it Hitomi} observations of the ICM~\citep{Hitomi2018temp} show that the turbulent motions in the Perseus cluster of galaxies have $\delta u \sim 160$~km/s, consistent with sub-sonic turbulence and an Alfv\'enic Mach number ${\rm M}_{\rm A}\gtrsim 1$, given typical values of the plasma $\beta \sim 100$ implied by Faraday rotation measurements of the intracluster magnetic field in several clusters~\cite{Murgia2004,Vogt2005,Bonafede2010,Govoni2017,Stuardi2021}. Similarly, the turbulence in black-hole accretion flows is expected to be composed of incompressible fluctuations, as indicated by local shearing-box simulations of the magnetorotational instability \citep{WalkerBoldyrevLesur2016,Kunz2016,Bacchini2022} (although a recent study shows a comparable amount of slow-mode fluctuations~\cite{Kawazura2021}).

We performed multiple simulations of driven turbulence in high-$\beta$, collisionless plasmas. All simulations employ an elongated computational domain spanning a size of $L_x\times L_y\times L_z \approx (120.5 \rho_{\rm i0})^2\times 241\rho_{\rm i0}$ with $384^2\times768$ grid cells and 1000 macro-particles per cell. These dimensions imply perpendicular wavenumbers $k_\perp$ that span both an inertial range, with $k_{\perp,{\rm min}} \rho_{\rm i0} \approx 0.05$, and a kinetic (sub-Larmor) range, with $k_{\perp,{\rm max}} \rho_{\rm i0} = 10$. While it is not computationally feasible to run even larger simulations, we have tested the convergence with smaller simulations and obtained qualitatively similar results. The majority of the results presented in this paper (the exception being Figures~\ref{fig:track} and \ref{fig:collisionality}) are drawn from two simulations that have $\beta_{\rm i0} = 4$ and $16$, but identical energy injection rates per volume $\varepsilon_{\rm dr} = n_{\rm i0} (L_x/L_z)^2 v_{\rm A0}^2/2 t_{\rm corr}$. The latter is expected to drive critically balanced fluctuations at the outer scale of the box with amplitudes $\delta u_L/v_{\rm A0} \sim L_x/L_z = 0.5$; the actual strength of the fluctuations is time-dependent and can be different from this value, depending on the response of the plasma to the driving (the values measured in our runs are ${\rm M}_{\rm A} \approx 0.35$ for $\beta_{\rm i0} = 16$ and ${\rm M}_{\rm A} \approx 0.48$ for $\beta_{\rm i0} = 4$). In these two simulations, the forcing excites fluctuations with $k_\perp \in [1,2] k_{\perp,{\rm min}}$ and $k_\parallel \in [1,2] k_{\parallel,\rm min}$. We have also tested driving with $k_\perp = k_{\perp,{\rm min}}$ and $k_\parallel = k_{\parallel,{\rm min}}$ in an additional run, which produced more coherent large-scale modes than did the other forcing scheme (see Figure~\ref{fig:track}). We have also reproduced our results qualitatively and quantitatively using smaller runs with $\delta u_L/v_{\rm A0} \sim L_x/L_z = L_y/L_z = 1.0$. Each simulation was run for several Alfv\'en crossing times $t_{\rm cross} \equiv L_\parallel/v_{\rm A0}$ to achieve quasi-steady state (these simulations are never in a true steady state because of continued energy injection from driving and, consequently, continued heating of the underlying ion distribution). For our $\beta_{\rm i0} = 16$ simulations, $t_{\rm cross} \approx 964 \Omega_{\rm i0}^{-1}$; for $\beta_{\rm i0} = 4$ ones, $t_{\rm cross} \approx 482 \Omega_{\rm i0}^{-1}$.

Most of the energy injected by forcing leads to particle energization. The remaining energy cascades below the ion Larmor scale, which can eventually lead to a pile-up of magnetic energy at the grid scale (because we employ hybrid-kinetic model, there are no electron-kinetic and -inertial effects to absorb this energy or unfreeze the magnetic flux). To mitigate this effect, we add hyper-resistive dissipation with the value of resistivity just large enough to dissipate energy at the grid scale. This hyper-resistivity is implemented as an additional electric-field component in~\eqref{eq:induction} of the form $\bb{E}_{\rm hyper} = -\eta_{\rm hyper} \grad^2 \bb{J}$, where $\bb{J} = (c/4\pi)\grad \btimes \bb{B}$ is the electric current.

\subsection{Evolution of high-$\bb{\beta}$ turbulence}
\label{sec:disruption}

In this Section, we summarize the evolution of turbulent fluctuations in our simulations of high-$\beta$ turbulence. This evolution is illustrated with three-dimensional snapshots of the magnetic field in Figure~\ref{fig:3D} and is separated into several stages: the excitation of mirror and AIC instabilities at early times, when the external driving dominates; an intermediate steady state dominated by Landau damping; and a quasi-steady state characterized by the co-existence of a turbulent cascade and micro-fluctuations associated with the firehose instability.

\subsubsection{Excitation of mirror and ion-cyclotron instabilities}
\label{sec:mirror-aic-stage}

\begin{figure}
    \centering
    \includegraphics[width=\columnwidth]{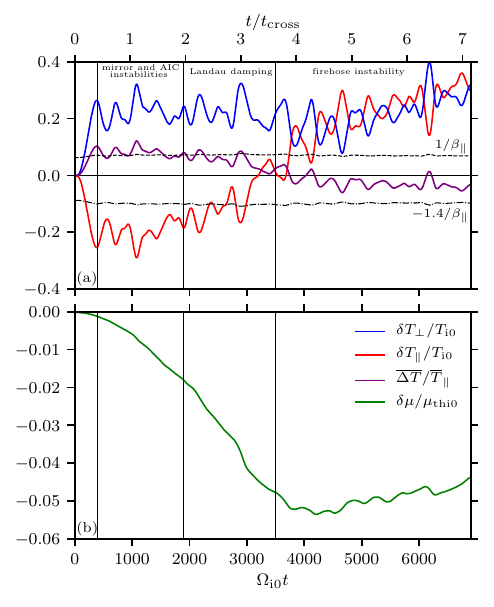}
    \caption{(a) Time-evolution of box-averaged perpendicular and parallel temperatures of the plasma (${\delta T_{\perp,\parallel}} \equiv \overline{T}_{\perp,\parallel} - T_{\rm i0}$), and box-averaged temperature anisotropy $\overline{\Delta T} = {\delta T}_\perp - {\delta T}_{\parallel}$ in $\beta_{\rm i0} = 16$ simulation. Different stages of the simulation are marked with vertical lines. See text for more details about individual stages: \S\ref{sec:mirror-aic-stage} and Appendix~\ref{sec:appendix_instabilities} for mirror and AIC stage, \S\ref{sec:landau-stage} for Landau-damping stage, \S\ref{sec:firehose-stage} for quasi-steady-state firehose stage. (b) Evolution of box-averaged magnetic moment of particles ($\delta \mu \equiv \overline{\mu} - \mu_{\rm thi0}$). During the mirror, AIC, and Landau-damping stages, the average magnetic moment decreases due to scattering off mirror and AIC fluctuations. In the quasi-steady state, it slowly increases, because of ion heating and coupling of perpendicular and parallel temperatures due to scattering of ions off firehose fluctuations.}
    \label{fig:T_vs_t}
\end{figure}

In our simulations of high-$\beta$ turbulence, large-scale fluctuations in the bulk ion velocity are driven by the external forcing. These motions cause the volume-averaged magnetic-field strength to increase, which produces positive temperature anisotropy, $\Delta T \equiv T_\perp - T_\parallel > 0$, through adiabatic invariance. Figure~\ref{fig:T_vs_t} shows the evolution of the box-averaged parallel (red line) and perpendicular (blue line) temperatures during the simulation at $\beta_{\rm i0} = 16$. (Density fluctuations in our simulation are small, so adopting density-weighted averaging does not change the results significantly.) The box-averaged temperature anisotropy, $\overline{\Delta T} = {\delta T_\perp} - {\delta T_\parallel}$, where ${\delta T_{\perp,\parallel}} \equiv \overline{T}_{\perp,\parallel} - T_{\rm i0}$, is shown with the purple line. Very early in the simulation, the box-averaged temperature anisotropy, $\overline{\Delta T}/\overline{T}_{\parallel}$, reaches the $1/\beta_\parallel$ threshold for the mirror instability. Some parts of the box also cross the threshold for rapid growth of the AIC instability, which we take to be $0.5/\sqrt{\beta_\parallel}$ following Refs.~\citep{SagdeevShafranov1960,Gary1994}. (The AIC instability is technically threshold-less, but its growth rate decreases exponentially for pressure anisotropies below ${\approx}0.5/\sqrt{\beta_\parallel}$; see Refs.~\citep{Gary1992, Hellinger2006,Sironi2015} for additional details on the AIC instability's thresholds.) As a result, mirror and AIC fluctuations appear at small scales. These fluctuations can be seen in the three-dimensional snapshot of $\delta B_\parallel$ in Figure~\ref{fig:3D} (left column), and in more detail in Figure~\ref{fig:slice_mirror} in Appendix~\ref{sec:appendix_instabilities}, where we also present a detailed structure-function analysis of mirror and AIC modes. Ultimately (after ${\approx}$2--3 Alfv\'en-crossing times), the temperature anisotropy falls below the mirror threshold and positive-pressure-anisotropy instabilities are no longer driven, meaning that the mirror and AIC stages are just transient in our simulations. In the quasi-steady state, only a small portion of the simulation box is above the mirror threshold, and mirror fluctuations are not obviously present (see Figure~\ref{fig:anisotropy} for more detail).

\subsubsection{Landau-damping stage}
\label{sec:landau-stage}

\begin{figure}
    \centering
    \includegraphics[width=\columnwidth]{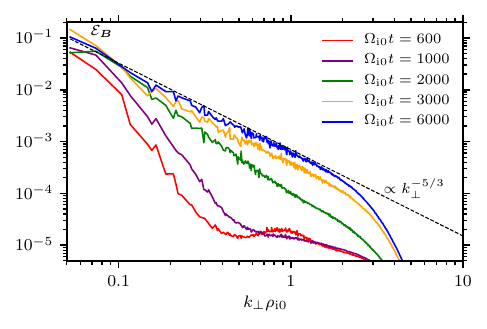}
    \caption{Time-evolution of the magnetic-energy spectra in the $\beta_{\rm i0} = 16$ simulation. In the initial phase of the simulation (red line, \S\ref{sec:mirror-aic-stage}), mirror instability is triggered, and the spectrum has a bump at kinetic scales. In the Landau-damping stage (green and orange lines, \S\ref{sec:landau-stage}), the spectrum is steeper than $k_\perp^{-5/3}$ due to dissipation of turbulence. In the quasi-steady state (blue line, \S\ref{sec:firehose-stage}), firehose fluctuations are produced, and the spectrum becomes slightly shallower than $k_\perp^{-5/3}$.}
    \label{fig:spectrum_vs_t}
\end{figure}

After this initial transient, the simulation reaches an intermediate quasi-steady state. Its key characteristic is growth of the parallel temperature of the plasma. As we show in Figure~\ref{fig:T_vs_t}, at the start of the $\beta_{\rm i0} = 16$ simulation (before $t\approx0.3t_{\rm cross}$), the temperatures evolve adiabatically: conservation of particles' adiabatic invariants implies $T_\perp \propto B$ and $T_\parallel \propto B^{-2}$, and thus the temperature anisotropy is driven towards positive values as $B$ increases. Once $\Delta T/T_\parallel$ reaches the $1/\beta_{\parallel}$ threshold of the mirror instability, Larmor-scale magnetic mirrors are produced, which limit further growth of pressure anisotropy by trapping particles in the deepening troughs of the mirrors where the total magnetic-field strength is approximately constant \citep{Kunz2014,Rincon2015,Melville2016}. 

Further evolution of the system is influenced by two processes. Pitch-angle scattering off the AIC fluctuations and the edges of strong mirror fluctuations reduces the average magnetic moments of particles (Figure~\ref{fig:T_vs_t}b). At the same time, the parallel temperature steadily grows during this period. Part of this growth can be attributed to pitch-angle scattering, occurring at a rate that can be estimated from the evolution of the average magnetic moment (Figure~\ref{fig:T_vs_t}b). From $\Omega_{\rm i0} t=2000$ to $3500$ (or from ${\sim}2$ to ${\sim}3.5$ $t_{\rm cross}$), it changes by $\delta \mu/\mu_{\rm thi0} \sim 0.035$, corresponding to $\delta T_\perp/T_{\rm i0}\sim 0.035$. Assuming that this change is due to pitch-angle scattering, $\delta T_\parallel/T_{\rm i0} \sim 2\delta T_\perp/T_{\rm i0} \sim 0.07$, which is only enough to explain ${\sim}25\%$ of the parallel heating during the same time period. We interpret the remaining energization as caused by the Landau damping of Alfv\'enic fluctuations. This interpretation is also supported by the spectrum of magnetic energy ($\mathcal{E}_{\bs{B}}$) being much steeper than $-5/3$ (see Figure~\ref{fig:spectrum_vs_t}), in agreement with gyrokinetic results~\citep{Kawazura2018}. In contrast, in kinetic simulations of $\beta\lesssim 1$ turbulence \citep{Arzamasskiy2019,Cerri2021}, a spectrum with a slope of $-5/3$ develops within ${\sim}1$--$2$ Alfv\'en crossing times. Additional evidence for Landau damping can be found in the evolution of the ion distribution function, which shows flattening near the  Alfv\'en speed (see Figure~\ref{fig:dist}) and from analysis of the field-particle correlation function (not included in the paper), which exhibits resonant features near $v_{\rm A}$, consistent with expectations from Landau damping~\cite{Howes2017}. Details of this diagnostic can be found in Refs.~\cite{Arzamasskiy2019,Cerri2021}.

This Landau-damping phase continues until the pressure anisotropy becomes negative, the firehose instability is triggered, and the simulated turbulence reaches a quasi-steady state. A very rough estimate for the time required for Landau damping to drive the pressure anisotropy beyond the firehose threshold may be obtained by supposing that the entire cascade rate were dissipated as parallel energization. In this case, the parallel temperature would grow large enough to produce a firehose-unstable pressure anisotropy within a time $t/t_{\rm cross} \sim n \Delta T_\parallel \varepsilon_{\rm dr}^{-1} t_{\rm cross}^{-1} \sim \Delta p_\parallel n^{-1} {\rm M}_{\rm A}^{-2} v_{\rm A}^{-2} \sim v_{\rm th}^2 \beta^{-1} {\rm M}_{\rm A}^{-2} v_{\rm A}^{-2} = {\rm M}_{\rm A}^{-2}$. 
For our simulations having ${\rm M}_{\rm A}\approx 0.5$, this time is ${\sim}4 t_{\rm cross}$. 

\subsubsection{Quasi-steady state}
\label{sec:firehose-stage}

\begin{figure}
    \centering
    \includegraphics[width=\columnwidth]{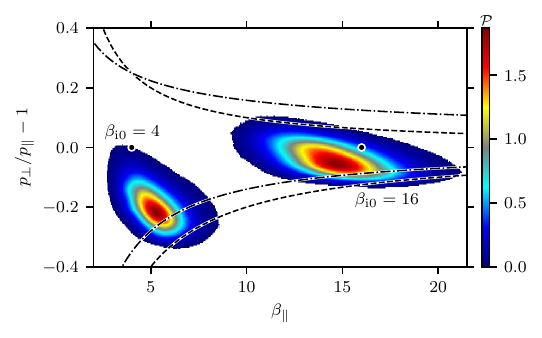}
    \caption{A probability density function of pressure anisotropy and parallel plasma beta for simulations with $\beta_{\rm i0}=4$ and $\beta_{\rm i0} = 16$. Histograms are normalized so that the integral over $\beta_\parallel$ and $p_\perp/p_\parallel - 1$ equals $1$. Black dots indicate the positions of these simulations at the start of each run. Dashed and dot-dashed lines represent the threshold of various kinetic micro-instabilities (see text for more detail). Despite being initially driven towards positive values, the quasi-steady-state pressure anisotropy is negative, and is close to the $-1.4/\beta_{\parallel}$ threshold of kinetic firehose instability (dot-dashed line).}
    \label{fig:anisotropy}
\end{figure}

Once the combination of pitch-angle scattering and Landau damping drives the pressure anisotropy beyond the firehose threshold, magnetic-field fluctuations grow on ion-Larmor scales at the expense of the anisotropy in the distribution function. A quasi-steady state results in which the magnetic spectrum acquires power at small scales that locally flattens it to be shallower than $k^{-5/3}_\perp$ and the box-averaged pressure anisotropy is close to zero but slightly negative (see Figures~\ref{fig:T_vs_t} and \ref{fig:spectrum_vs_t}, respectively). Figure~\ref{fig:anisotropy} provides further information on the pressure anisotropy in the quasi-steady state by showing its distribution versus $\beta_\parallel$ (so-called ``Brazil'' plots, extensively used in the solar-wind community~\cite{Kasper2002,Bale2009}). The accompanying dashed lines indicate the thresholds of the mirror instability ($1/\beta_\parallel$) and of the fluid firehose instability ($-2/\beta_\parallel$), beyond which the Alfv\'en speed becomes imaginary. The dot-dashed line at positive pressure anisotropy represents the threshold of the AIC instability (${\approx}0.5/\sqrt{\beta_\parallel}$), which is active in the beginning of the run. At negative pressure anisotropy, the dot-dashed line shows the approximate threshold (${\approx}-1.4/\beta_{\parallel}$) of the kinetic firehose instability \citep{Bott2021,Bott2022}. Unlike in the mirror and AIC snapshots (Figures~\ref{fig:slice_mirror} and \ref{fig:slice_IC} of Appendix~\ref{sec:appendix_instabilities}), the pressure anisotropy is bound between the mirror and firehose instability thresholds, hugging the latter more closely. The average pressure anisotropy is negative, and its absolute value is smaller than $1/\beta$. 

This distribution of pressure anisotropy in the quasi-steady state is very different from what has been found in comparable Braginskii-MHD simulations of Alfv\'{e}nic guide-field turbulence~\citep{Squire2019} and of magnetorotational turbulence~\citep{Kempski2019}, in which a significant amount of the simulated plasma sits up against either the mirror or the firehose threshold \footnote{See also Ref.~\citep{Squire2023} for the more recent Braginskii-MHD simulations of high-$\beta$ turbulence.}. There are (at least) two reasons for this difference. First, our kinetic simulations allow for collisionless damping, which we find constantly pushes the simulated plasma toward the firehose threshold by parallel heating the ions. Second, once the kinetic instabilities are triggered, they self-consistently produce Larmor-scale mirror and firehose fluctuations, which regulate the pressure anisotropy. These fluctuations decay rather slowly \citep{Melville2016,Kunz2020}, and continue to scatter particles towards isotropy even after the pressure anisotropy has returned below the instability thresholds. As a result, $\Delta p$ is reduced much more efficiently and its mean quasi-steady-state value is smaller than seen in Braginskii-MHD simulations, which account for these instabilities with a large ``limiter'' collisionality that is active only when the plasma ventures beyond the stability thresholds.

\begin{figure*}
    \centering
    \includegraphics[width=\textwidth]{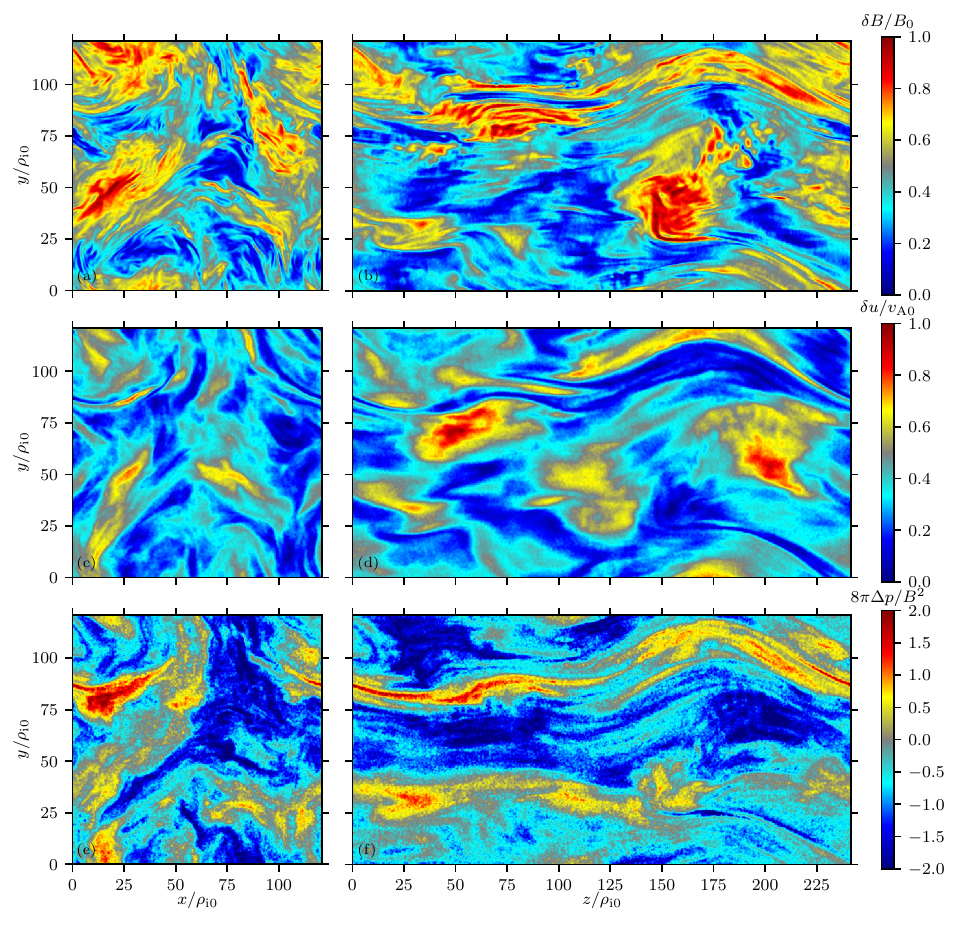}
    \caption{Snapshots of the magnetic-field-strength fluctuations (a,b), fluctuations of the flow velocity (c,d), and of pressure anisotropy normalized to the magnetic pressure (e,f), all taken in the quasi-steady state of $\beta_{\rm i0} = 16$ simulation. There is much more small-scale structure in the magnetic field relative to the flow velocity, which indicates that the effective magnetic Prandtl number is large. Mirror and AIC fluctuations are not apparent, unlike in the earlier stages of the same simulation (Figures~\ref{fig:slice_mirror} and ~\ref{fig:slice_IC}). Instead, there are small-scale oblique modes in the regions with large negative pressure anisotropy (e.g., near $z\approx 25 \rho_{\rm i0}, y\approx 75 \rho_{\rm i0}$ and $z\approx 130 \rho_{\rm i0}, y\approx 10 \rho_{\rm i0}$). We associate these fluctuations with the firehose instability.}
    \label{fig:slice_steady_state}
\end{figure*}

Figure~\ref{fig:slice_steady_state} shows slices of the magnetic-field strength, plasma fluid velocity, and pressure anisotropy normalized to magnetic pressure, $8\pi\Delta p/B^2$, all in the quasi-steady state of the $\beta_{\rm i0} = 16$ simulation. Mirror fluctuations, manifest at the beginning of the run (Figure~\ref{fig:slice_mirror}), are no longer visible. This is likely due to their decay and shearing by the fluctuations associated with the turbulent cascade. One can still see some quasi-parallel fluctuations in regions with positive pressure anisotropy (e.g., near $z\approx 25 \rho_{\rm i0}, y\approx 75 \rho_{\rm i0}$ and $z\approx 130 \rho_{\rm i0}, y\approx 10 \rho_{\rm i0}$), which we attribute to AIC instability. Additionally, one can notice small-scale oblique modes of large amplitude (such as those at $y\approx 75 \rho_{\rm i0}$ and $z\approx 200 \rho_{\rm i0}$), which are correlated with regions of negative pressure anisotropy. Studies of localized firehose instability~\cite{Kunz2014} show that this instability produces similar oblique modes, and so we interpret those fluctuations as firehose modes produced continually in the quasi-steady state. 

In the quasi-steady state, the box-averaged magnetic moment of the particles slowly increases with time (Figure~\ref{fig:T_vs_t}b). The reason for this is that the perpendicular and parallel components of the temperature are coupled to one another via pressure-anisotropy regulation by the firehose instability (although many different instabilities are present in our simulations, the firehose is much more efficient at scattering particles than mirror \citep{Kunz2014}, so the temperatures are well coupled only after firehose is triggered). As a result, as turbulence is dissipated into particle heat, those components increase together, which causes slow growth of the average magnetic moment. In our simulation setup, kinetic energy is constantly injected into the box, so we expect such an increase in average magnetic moment to continue indefinitely. The changes of the magnetic moments of some individual particles are much more rapid, as we discuss in~\S\ref{sec:collisionality}.

\subsection{Energy transfer in high-$\bb{\beta}$ turbulence}
\label{sec:energy_transfer}

The results described in the previous subsections suggest that collisionless high-$\beta$ turbulence contains a superposition of local interactions (a Kolmogorov-like cascade from large to small scales) and non-local processes mediated by kinetic micro-instabilities (mirror and AIC during the driving stage, and firehose in the quasi-steady state). In this Section, we explore the relative importance of different energy-transfer channels. To do that, we calculate various ``transfer functions'' that diagnose quantitatively the scale-to-scale transfer of energy in the turbulence caused by different bulk forces \cite{Kraichnan1967,Grete2017}. A detailed derivation of these diagnostics is given in Appendix~\ref{sec:appendix_tf}. In brief, the equations for the bulk kinetic, thermal, and magnetic energies are written in a form that makes explicit the transfer of energy between different ``reservoirs'' of energy and different wavenumbers. These reservoirs are defined as vector fields $\bb{a}$ with associated energy $\bb{a}^2/2$. For the bulk kinetic energy, the corresponding vector field is $\bb{a}^u \equiv \sqrt{\varrho}\bb{u}$; for the magnetic energy it is $\bb{a}^B \equiv\bb{B}/\sqrt{4\pi}$. For the thermal energy, multiple definitions are possible. In this paper, we choose $\bb{a}^{\Delta p} \equiv \sqrt{|\Delta p|} \hat{\bb{b}}$, so that the viscous stress in the momentum equation may be written as $\bb{a}^{\Delta p} \bb{a}^{\Delta p}$, similar to the Maxwell stress. Only the part of the thermal energy associated with $\Delta$ (``anisotropic thermal energy'') is included in our definition of $\bb{a}^{\Delta p}$; the remaining part of thermal energy, ``isotropic thermal energy,'' is equal to $3p_\perp/2$. Our definition of $\bb{a}^{\Delta p}$ is also consistent with the expression for the free energy in the ``kinetic reduced MHD'' limit (i.e., the long-wavelength limit of gyrokinetics)~\citep{Kunz2015},
\begin{align}
    W_{\rm KRMHD}^{\Delta} & = \frac{B_0^2}{8\pi} \int \beta_{\parallel, {\rm i}} \frac{\Delta_{\rm i}}{2}\frac{\delta B_\perp^2}{B_0^2}\,{\rm d}^3\bb{x} \nonumber \\
    & = \frac{B_0^2}{8\pi}  \int \vartheta_{\Delta p}\frac{(\delta \bb{a}^{\Delta p})^2}{2}\,{\rm d}^3\bb{x}, \label{eq:W_krmhd}
\end{align}
where $\vartheta_{\Delta p}\equiv {\rm sign}(\Delta p)$, $\delta \hat{\bb{b}} = \delta \bb{B}_\perp/B_0 \ll 1$, and the pressure anisotropy is assumed to have an absolute value much larger than its fluctuation due to $\delta \bb{B}$ [viz., $\Delta_{\rm i} = \overline{\Delta_{\rm i}} + \mathcal{O} (|\delta \hat{\bb{b}}|^2)$, which allows us to neglect terms proportional to $\delta\Delta_{\rm i}^2$ and $\delta \Delta_{\rm i} \delta B_\perp$ and write $\delta \bb{a}^{\Delta p}  = \sqrt{|\Delta p|} \delta \hat{\bb{b}}$]. The definition of $\bb{a}^{\Delta p}$ contains the absolute value of the pressure anisotropy. Such a choice was made from a purely technical standpoint, in order to avoid imaginary values when computing this vector at each point of the simulation domain. This choice also forces us to include the sign of $\Delta p$ in various expressions, such as in equation~(\ref{eq:W_krmhd}), which can make results hard to interpret if $\vartheta_{\Delta p}$ changes significantly throughout the domain. Fortunately, as we show in Figure~\ref{fig:anisotropy}, most of the domain has negative pressure anisotropy, and the volume fraction with positive $\Delta p$ is small. 

Given the definitions of the energy ``reservoirs,'' some terms in the equation for the bulk kinetic energy can be written in the form of shell-to-shell transfer functions. The most important transfer functions in that equation are the transfer function due to the Reynolds stress,
\begin{equation}
    \mathcal{T}^{\rm R}_{q_\perp \rightarrow k_\perp} \equiv -\int \langle \bb{a}^u\rangle_{k_\perp} \bcdot \bb{u} \bcdot \grad \langle \bb{a}^u\rangle_{q_\perp}\,{\rm d}^3 \bb{x},\label{eq:tf_R}
\end{equation}
the transfer function due to the Maxwell stress,
\begin{equation}
    \mathcal{T}^{\rm M}_{q_\perp \rightarrow k_\perp} \equiv \int \langle \bb{a}^u\rangle_{k_\perp} \bcdot \frac{\bb{B}}{\sqrt{4\pi\varrho}} \bcdot \grad \langle \bb{a}^B\rangle_{q_\perp}\,{\rm d}^3 \bb{x},
\end{equation}
and the transfer function due to the anisotropic viscous stress,
\begin{equation}
    \mathcal{T}^{\rm V}_{q_\perp \rightarrow k_\perp} \equiv \int \langle \bb{a}^u\rangle_{k_\perp} \bcdot \vartheta_{\Delta p} \sqrt{\frac{|\Delta p|}{\varrho}}\hat{\bb{b}} \bcdot \grad \langle \bb{a}^{\Delta p}\rangle_{q_\perp}\,{\rm d}^3 \bb{x}.\label{eq:tf_V}
\end{equation}
Each of these transfer functions represents the rate of energy transfer between fluctuations whose wavenumbers lie within the logarithmic perpendicular-wavenumber shells centered on $q_\perp$ and $k_\perp$ (from kinetic to kinetic in the case of $\mathcal{T}^{\rm R}$, from magnetic to kinetic for $\mathcal{T}^{\rm M}$, and from anisotropic thermal energy to kinetic for $\mathcal{T}^{\rm V}$).

\begin{figure*}
    \centering
    \includegraphics[width=\textwidth]{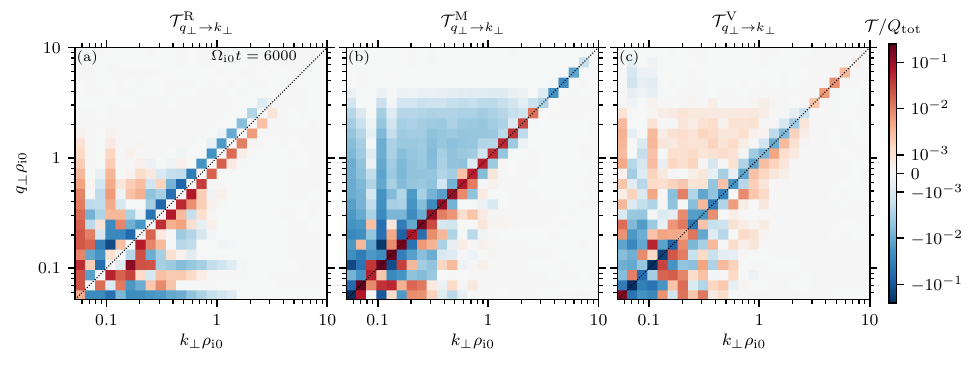}
    \caption{Energy transfer functions (\ref{eq:tf_R})--(\ref{eq:tf_V}) due to the Reynolds (left), Maxwell (middle), and viscous stresses (right) in the quasi-steady state of our $\beta_{\rm i0} = 16$ simulation. There is a considerable amount of non-local energy transfer due to the firehose instability, which is not present during the early stages of the simulation (see Appendix \ref{sec:appendix_tf} for an analogous plot featuring the earlier stages).}
    \label{fig:tf_2d}
\end{figure*}

Figure~\ref{fig:tf_2d} shows the flow of kinetic energy through Fourier space as calculated by the transfer functions (\ref{eq:tf_R})--(\ref{eq:tf_V}) in the two-dimensional plane of wavenumbers $q_\perp$ and $k_\perp$ in quasi-steady state of $\beta_{\rm i0} = 16$ simulation at $\Omega_{\rm i0} t = 6000$. Results from the same diagnostic obtained during earlier stages of this simulation can be found in Appendix~\ref{sec:appendix_tf}, Figure~\ref{fig:nonlocal}. The sign of the transfer terms in Figure~\ref{fig:tf_2d} represents the change in bulk kinetic energy in shell $k_\perp$. In the early stages of the simulation, including the mirror/AIC and the Landau-damping stages, the transfers are mostly local, and are similar to those found previously in simulations of MHD turbulence~\citep{Grete2017}. In contrast, in the quasi-steady state, there is considerable non-local energy transfer associated with the Maxwell and viscous stresses (see the $q_\perp > k_\perp$ part of Figure~\ref{fig:tf_2d}b,c). This non-local transfer takes energy from the large-scale fluid motions and transfers it into small-scale magnetic fields, as is expected from the firehose instability. We thus conclude that kinetic micro-instabilities are active even in the quasi-steady state and that they contribute a non-negligible amount of energy transfer. Out of the three instabilities that we see in our simulations, the one most responsible for the non-local energy transfer appears to be the firehose, as this non-local transfer becomes competitive with the local transfers only after the box-averaged pressure anisotropy approaches the firehose threshold and localized patches of the plasma exceed that threshold. The viscous stress in the quasi-steady state has both local and non-local components, which we discuss in more detail in~\S\ref{sec:heating}, and is mostly negative, indicating that the viscous stress mostly removes energy from the bulk motions. This transfer causes the conversion of energy between bulk kinetic and anisotropic thermal energies, and subsequently steepens the kinetic-energy spectrum (as we show in~\S\ref{sec:viscosity}). 

The energy transfer quantified by the transfer functions (\ref{eq:tf_R})--(\ref{eq:tf_V}) is, in principle, reversible. Indeed, for any energy transfer term of the form $\bb{a}_1 \bcdot \bb{f} \bcdot\grad \bb{a}_2$ for some vector field $\bb{f}$, there is another term of the form $\bb{a}_2 \bcdot \bb{f} \bcdot\grad \bb{a}_1$ (see Appendix~\ref{sec:appendix_tf} for the transfer terms in the induction equation and in the equation for $\Delta p$). The total energy transfer due to such terms,
\begin{align}\label{eq:tf_conservation_law}
    & \sum_{k_\perp,q_\perp} \mathcal{T}_{q_\perp \rightarrow k_\perp}^{a_1\rightarrow a_2}+ \mathcal{T}_{q_\perp \rightarrow k_\perp}^{a_2\rightarrow a_1}  \\*
    & = \int \left(\bb{a}_1 \bcdot \bb{f} \bcdot \grad \bb{a}_2 + \bb{a}_2 \bcdot \bb{f} \bcdot \grad \bb{a}_1\right)\,{\rm d}^3 \bb{x}  \nonumber \\*
    & = \int \bb{f} \bcdot \grad \left(\bb{a}_1\bcdot\bb{a}_2\right)\,{\rm d}^3 \bb{x} = - \int \left(\bb{a}_1\bcdot\bb{a}_2\right) \grad\bcdot \bb{f}\,{\rm d}^3 \bb{x}, \nonumber
\end{align}
is zero if $\grad \bcdot \bb{f}$ is zero, which is one of the assumption we are making in our transfer-function analysis. We check this assumption {\it a posteriori} in Appendix~\ref{sec:appendix_tf} by computing ``compressive'' terms and comparing them to ``advection-like'' terms such as (\ref{eq:tf_R})--(\ref{eq:tf_V}). Although equation~\eqref{eq:tf_conservation_law} is valid for arbitrary $\bb{f}$, this vector in practice is proportional to $\bb{u}$ or $\bb{B}$. We refer the reader to Appendix~\ref{sec:appendix_tf}, and Figure~\ref{fig:nonlocal} in particular, for more information. One of the terms, which we do not consider explicitly, is the transfer of energy due to an effective collisionality [i.e., the last term in equation \eqref{eqn:cgl}]. This term is quite important in the quasi-steady state, as it is responsible for the conversion of energy between the anisotropic ($\Delta p/2$) and isotropic ($3p_\perp/2$) thermal energies: it reduces the pressure anisotropy and leads to irreversible heating. It is not feasible to compute this term directly in our simulations because, unlike the last term of equation \eqref{eqn:cgl}, the effective collisionality in our runs comes from wave--particle interactions and its analysis requires computation of high-order moments of the distribution function. The only other irreversible term is resistive dissipation, which we find to be relatively less important.

\subsection{Effective viscous scale and sub-viscous turbulence}
\label{sec:viscosity}

\begin{figure*}
    \centering
    \includegraphics[width=\textwidth]{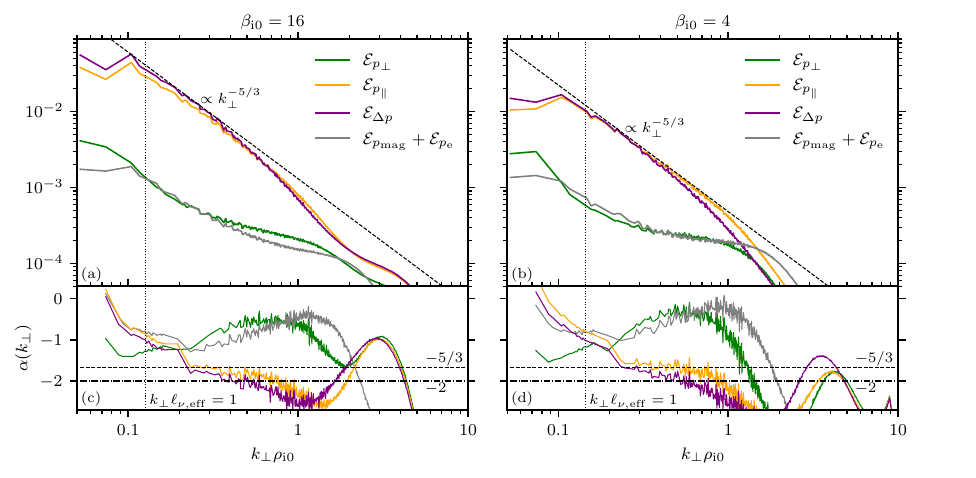}
    \caption{(a,b) Spectra of parallel (orange)  and perpendicular (green) pressures, pressure anisotropy (purple), and the sum of electron pressure and magnetic pressure (gray). We define the effective viscous scale $\ell_{\nu,{\rm eff}}$ as the scale at which the pressure anisotropy fluctuation amplitude ${\sim}k_\perp^{1/2} \Delta p_k$ peaks. This scale is close to the driving scale, in line with the analytical prediction of \S\ref{sec:theory}. (c,d) Spectral indices as a function of wavenumber for the spectra from upper panels.}
    \label{fig:spectrum_pressure}
\end{figure*}

\begin{figure*}
    \centering
    \includegraphics[width=\textwidth]{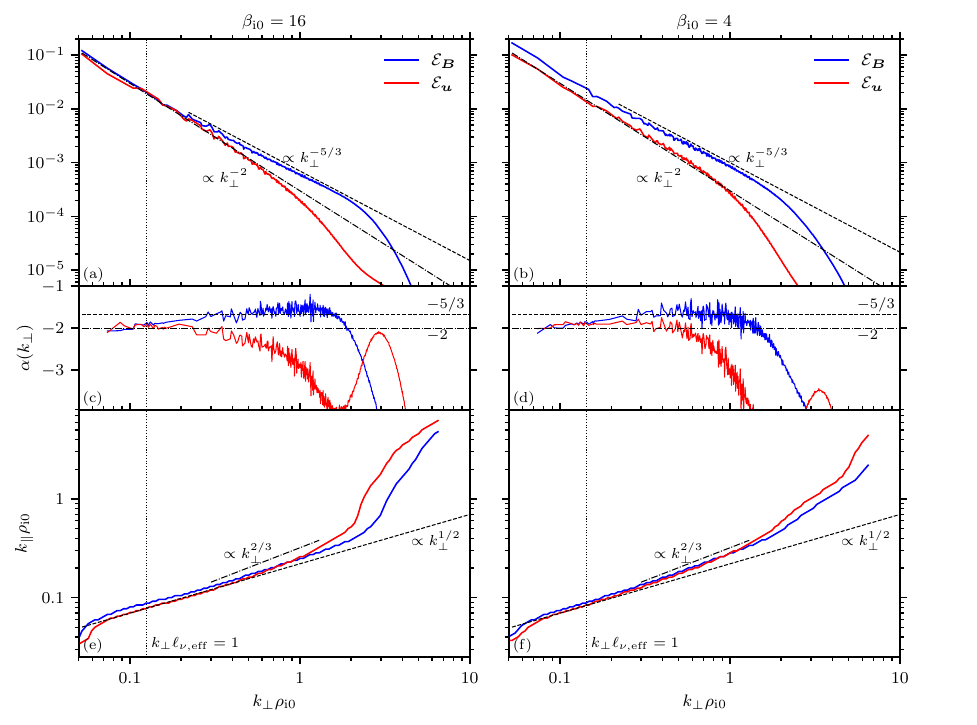}
    \caption{(a,b) Quasi-steady-state spectra of magnetic (blue) and kinetic (red) energies in the $\beta_{\rm i0} = 16$ (a) and $\beta_{\rm i0}=4$ (b) simulations. These spectra have similar slopes near the driving scale, but deviate from one another at scales below the effective viscous scale (defined in \S\ref{sec:viscosity} and in the caption of Figure~\ref{fig:spectrum_pressure}, and indicated by the dotted lines). The kinetic-energy spectra are steeper than $-5/3$ because of the anisotropic viscous stress; the slope of the  kinetic-energy spectrum becomes close to $-2$. The magnetic-energy spectrum becomes shallower towards $k_\perp \rho_{\rm i0} \sim 1$ because firehose fluctuations are injected at these scales in the quasi-steady state. (c,d) Spectral indices as functions of wavenumber for the magnetic- and kinetic-energy spectra. (e,f) Parallel wavenumber ($k_\parallel$) of the fluctuations as a function of their perpendicular wavenumber ($k_\perp$) obtained from magnetic-field structure-function analysis. The value of spectral anisotropy, $k_\parallel/k_\perp$, agrees with critical-balance predictions.}
    \label{fig:spectrum_steady_state}
\end{figure*}

In~\S\ref{sec:theory} we argued that, in critically balanced turbulence, the effective viscous scale associated with scattering by ion-Larmor-scale kinetic instabilities satisfies  $\ell_\nu \sim L {\rm M}_{\rm A}^{-3}$. This means that, for our simulations with ${\rm M}_{\rm A} \lesssim 1$, the viscous scale is expected to be close to, or, formally speaking, even above, the driving scale. Formally, we shall define the viscous scale as the scale at which the pressure anisotropy peaks. We obtain this scale in a way analogous to the regular energy-containing scale of the turbulence: if the velocity field has a spectrum of $\mathcal{E}_{\boldsymbol{u}} (k_\perp)$, then the energy-containing scale corresponds to the peak of $\delta u_k^2 \sim k_\perp \mathcal{E}_{\boldsymbol{u}} (k_\perp)$. For the viscosity, this means that the effective viscous scale $\ell_{\nu,{\rm eff}}$ corresponds to the value of $k_\perp$ at which the spectrum of pressure anisotropy $\mathcal{E}_{\Delta p}$ has a slope of $-1$. It is possible to define the effective viscous scale in other ways, e.g., by looking at the $k$-space peak of viscous dissipation. We do not use such a definition because of the non-local nature of the viscous dissipation (e.g., the spatial scale at which most of the kinetic energy is removed is not necessarily the same as the scale at which most of the thermal energy is deposited; see \S\ref{sec:energy_transfer}).

Figure~\ref{fig:spectrum_pressure} shows the spectra of the parallel and perpendicular pressures, as well as of the pressure anisotropy, in the quasi-steady state of the $\beta_{\rm i0} = 16$ and $\beta_{\rm i0} = 4$ simulations. The effective viscous scale for each simulation is indicated by a vertical dashed line: at $k_\perp \rho_{\rm i0} \approx 0.126$ for $\beta_{\rm i0} = 16$ and at $k_\perp \rho_{\rm i0} \approx 0.144$ for $\beta_{\rm i0} = 4$. These scales are close to the outer scale of the turbulence, consistent with the expectations presented in~\S\ref{sec:theory}. An interesting feature seen in Figure~\ref{fig:spectrum_pressure} is that the parallel-pressure spectrum is much larger in magnitude than the perpendicular-pressure spectrum. This difference may be explained as follows. If we assume perpendicular pressure balance, which is a natural outcome in high-$\beta$ anisotropic turbulence \citep{Schekochihin2009,Kunz2015}, and is also observed in our simulations (gray lines in Figure~\ref{fig:spectrum_pressure} correspond to the sum of electron pressure, $p_{\rm e} = n_{\rm e} T_{\rm e} = n_{\rm i} T_{\rm i0}$, and magnetic pressure, $p_{\rm mag} = B^2/8\pi$), that would imply $p_\perp + B^2/8\pi \approx {\rm const}$, and therefore $\delta p_\perp \sim -\delta B^2/8\pi$. At the same time, pressure anisotropy at sub-viscous scales (where the effective collisionality is less important \footnote{For a critically balanced cascade, the effective collisionality becomes unable to regulate pressure anisotropy when $\nu_{\rm eff}/k_\perp \delta u_\perp \sim 1$, which corresponds to $k_\perp L \sim {\rm M}_{\rm A}^{9/2} \beta^{3/2}$; for our $\beta_{\rm i0} = 16$ simulation, the latter is ${\approx}2.8$.}) behaves nearly adiabatically: ${\rm d}\Delta p/{\rm d}t \sim p\,{\rm d}\ln|B|/{\rm d}t$. This leads to pressure-anisotropy fluctuation of $\delta \Delta p \sim \beta\, \delta B^2 \gg \delta B^2 \sim \delta p_\perp$, and therefore $\delta p_\parallel \gg \delta p_\perp$. In the sub-viscous range, $\delta p_\parallel$ is passively advected, its spectrum has a similar slope to $\delta u_k$, and the parallel pressure remains larger than the perpendicular pressure throughout the sub-viscous range. Landau damping, viscous heating, and pitch-angle scattering off micro-fluctuations produced by kinetic micro-instabilities gives rise to large-scale fluctuations of $\delta p_\parallel$ to achieve a value of pressure anisotropy close to $\nu_{\rm eff}^{-1}\hat{\bb{b}}\hat{\bb{b}}\,\bb{:}\,\grad\bb{u}$.

To examine how the effective viscosity affects the cascade, we plot in Figure~\ref{fig:spectrum_steady_state} the quasi-steady-state spectra of magnetic energy (blue) and bulk kinetic energy (red) from both simulations. The magnetic-energy spectra have larger amplitudes than the kinetic-energy spectra at the driving scales due to changes in the effective Alfv\'en speed caused by the mean negative pressure anisotropy in the box (cf.~Figure~\ref{fig:anisotropy}). These spectra have similar slopes above the viscous scale. The kinetic-energy spectrum is steeper than $-5/3$ (close to $-2$), which is an indication of the viscous dissipation. Below the viscous scale, the kinetic-energy spectrum continues to steepen because of the sub-viscous ion heating (see~\S\ref{sec:heating} for more detail on ion heating). In contrast, the magnetic-energy spectrum becomes shallower. We attribute this flattening to small-scale energy injection by kinetic micro-instabilities \footnote{Another potential explanation could have been scale-dependence of the effective Alfv\'en speed. This explanation does not work for our simulations because pressure anisotropy is mostly concentrated at the outer scale.}. Given that the majority of the simulation box is near the threshold for the firehose instability (see Figure~\ref{fig:anisotropy}), we interpret this bump in the magnetic-energy spectrum as the spectrum of the firehose fluctuations.

\begin{figure}
    \centering
    \includegraphics[width=\columnwidth]{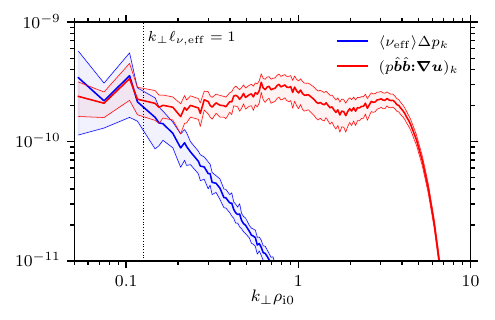}
    \caption{Comparison between the spectra of pressure anisotropy (blue) and $p\ROS$ (red), the latter of which is proportional to the rate of growth of the magnetic field [see equation \eqref{eqn:cgl}]. Above the effective viscous scale $\ell_{\nu,{\rm eff}}$ (vertical dotted line), the pressure anisotropy is proportional to $p\ROS$. The proportionality coefficient between the two (the ``Braginskii estimate'' for the effective collision frequency) is $\langle \nu_{\rm eff}\rangle \sim 0.01 \Omega_{\rm i0}$. At scales smaller than $\ell_{\nu,{\rm eff}}$, this effective collisionality is not large enough relative to the dynamical frequencies, so the pressure anisotropy deviates significantly from the Braginskii estimate. Note the bump in $\ROS$ near $k_\perp \rho_{\rm i0}\sim 1$, which corresponds to firehose fluctuations in quasi-steady state (the additional peak at $k_\perp \rho_{\rm i0}\approx 3$ is due to particle noise).}
    \label{fig:phase_shift}
\end{figure}

The deviation of the magnetic spectrum from the kinetic spectrum at sub-parallel-viscous scales indicates an important feature of high-$\beta$ turbulence: the small-scale firehose fluctuations are decoupled from the Alfv\'enic cascade. To illustrate this point, we plot in panels (e) and (f) of Figure~\ref{fig:spectrum_steady_state} the wavenumber anisotropy $k_\parallel(k_\perp)$ for $\delta \bb{u}$ and $\delta \bb{B}$ fluctuations. This dependence is computed using the values of the quasi-steady-state structure functions at $\bb{\ell} = \bb{\ell}_\perp$ and $\bb{\ell} = \bb{\ell}_\parallel$. The wavenumber anisotropy appears to have a scaling of $k_\parallel \propto k_\perp^{1/2}$ in the inertial range, consistent with critical balance:
\begin{equation}
    k_\parallel v_{\rm A} \sim k_\perp \delta u_\perp \propto k_\perp^{3/2} \mathcal{E}_{\boldsymbol{u}}^{1/2} \propto k_\perp^{1/2},
\end{equation}
where in the final step we have adopted a $k_\perp^{-2}$ spectrum. It is interesting that this critical balance is measured to hold at what are notionally sub-parallel-viscous scales, which suggests that the viscous backreaction on the cascade is not purely dissipative. The anisotropy computed using the magnetic-field fluctuations has the same scaling, despite the magnetic energy having a shallower spectrum. This means that the firehose modes, which are oblique and affect neither the parallel nor the perpendicular structure functions, do not participate in the critically balanced cascade. The magnetic spectrum is a superposition of the critically balanced cascade with a slope close to $-2$ and an additional spectrum of the firehose fluctuations, which peaks at kinetic scales. Independence of the firehose fluctuations from the Alfv\'{e}nic cascade has also been shown by Ref.~\citep{Bott2021} using an expanding box, in which the dominant contribution to the pressure anisotropy is from plasma expansion perpendicular to the mean field (rather than from the fluctuations).

\subsection{Effective collisionality}
\label{sec:collisionality}

In this Section, we estimate the effective collisionality from the results of our  hybrid-kinetic simulations. For that, we use two independent methods: one based on the evolution of the pressure stress in the simulations (\S\ref{sec:collisionality_1}) and one based on the motion of individual particles (\S\ref{sec:collisionality_2}).

\subsubsection{Effective collisionality from pressure-stress evolution}
\label{sec:collisionality_1}

In~\S\ref{sec:theory}, we employed a simple model \eqref{eqn:cgl}, where pressure anisotropy grows with the local parallel rate of strain $S \equiv \ROS$, and is relaxed by Coulomb collisions between particles at the rate $\nu$. A common assumption in reduced fluid models, such as Braginskii MHD, is that the typical frequency of fluctuations satisfies $\omega \ll \nu$, so ${\rm d}\Delta/{\rm d}t$ is much smaller than other terms in \eqref{eqn:cgl}. Then $\Delta \simeq S/\nu$. To test whether such a closure works in kinetic high-$\beta$ turbulence, we plot in Figure~\ref{fig:phase_shift} the spectra of $\Delta p$ (blue) and $p \ROS$ (red). The spectrum of pressure anisotropy is multiplied by a coefficient $\langle\nu_{\rm eff} \rangle \sim 0.01 \Omega_{\rm i0}$, which is the value of $(p \ROS)_{\bs{k}}/\Delta p_{\bs{k}}$ averaged over scales satisfying $k_\perp \ell_{\nu,{\rm eff}} < 1$. In what follows, we refer to the latter [which we also label as $(p \ROS)_L/\Delta p_L$] as the ``Braginskii estimate,'' because the pressure anisotropy in a weakly collisional plasma when the fluid motions are incompressible is given by Braginskii~\cite{Braginskii1965} as $\Delta p = p\ROS/\nu$. $\nu_{\rm eff}$ represents our approximation for the effective collisionality, which in our simulations is mediated by wave-particle interactions \footnote{A more detailed model for the effective collisionality will be presented in an upcoming publication by Bott et al.}. Shaded regions in Figure~\ref{fig:phase_shift} indicate root-mean-square fluctuations in pressure anisotropy and rate of strain measured during the quasi-steady state. The pressure-anisotropy spectrum follows the rate of strain down to the viscous scale, below which the pressure anisotropy starts to decrease while the parallel rate of strain increases. This increase corresponds to the injection of small-scale firehose fluctuations in quasi-steady state.

\begin{figure*}
    \centering
    \includegraphics[width=\textwidth]{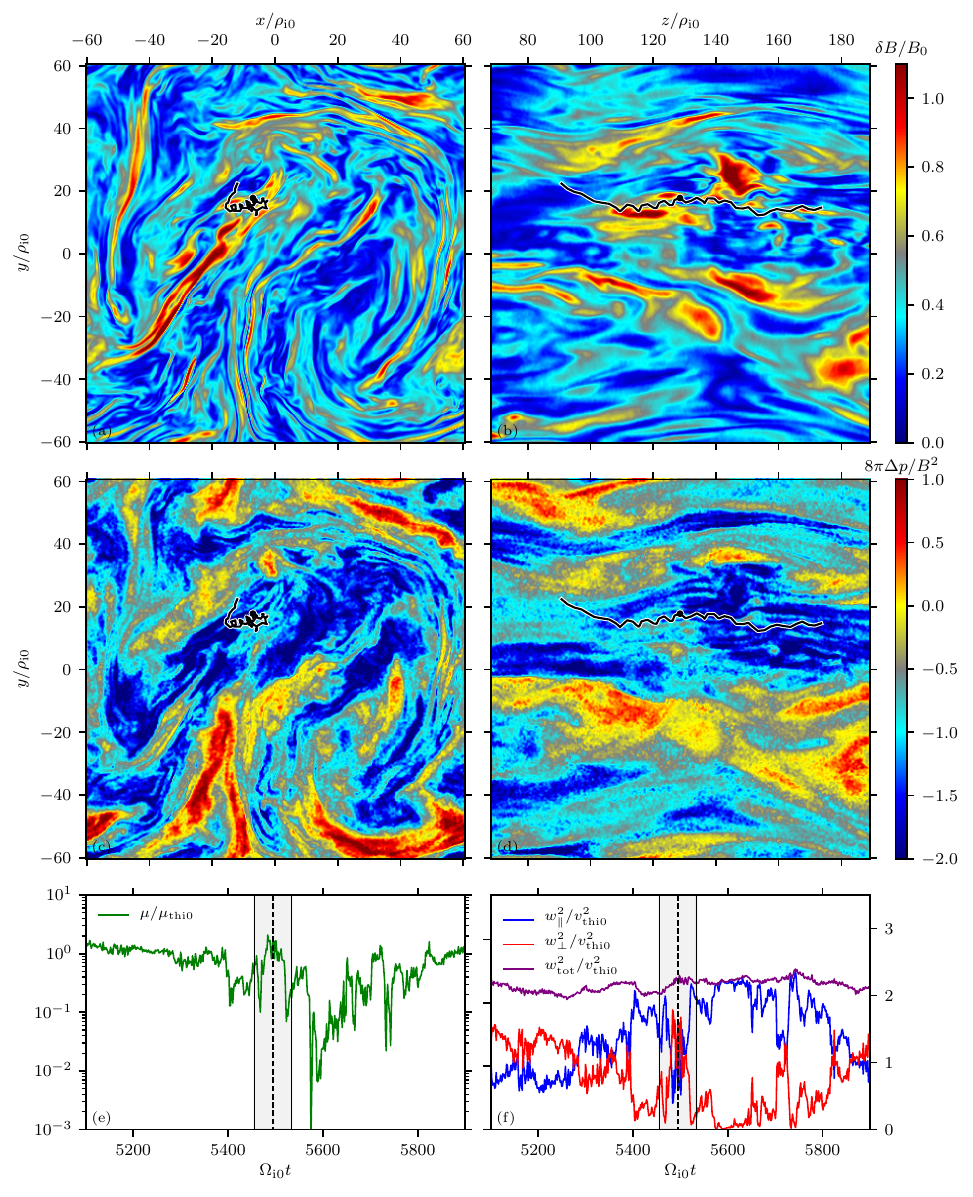}
    \caption{Trajectory of a tracked particle from the  $\beta_{\rm i0} = 16$ simulation with single-mode driving (see~\S\ref{sec:parameters}). This particle moves through a region with firehose-unstable pressure anisotropy (panels~c,d). The same region shows small-scale oblique magnetic-field fluctuations (panels~a,b), which we interpret as firehose modes. As the particle passes through the region, its magnetic moment starts to break (panel~e), and it experiences pitch-angle scattering at almost constant total energy (panel~f). The gray shaded regions in panels (e) and (f) indicate the period of time over which the trajectory in panels (a)--(d) is plotted.}
    \label{fig:track}
\end{figure*}

\begin{figure*}
    \centering
    \includegraphics[width=\textwidth]{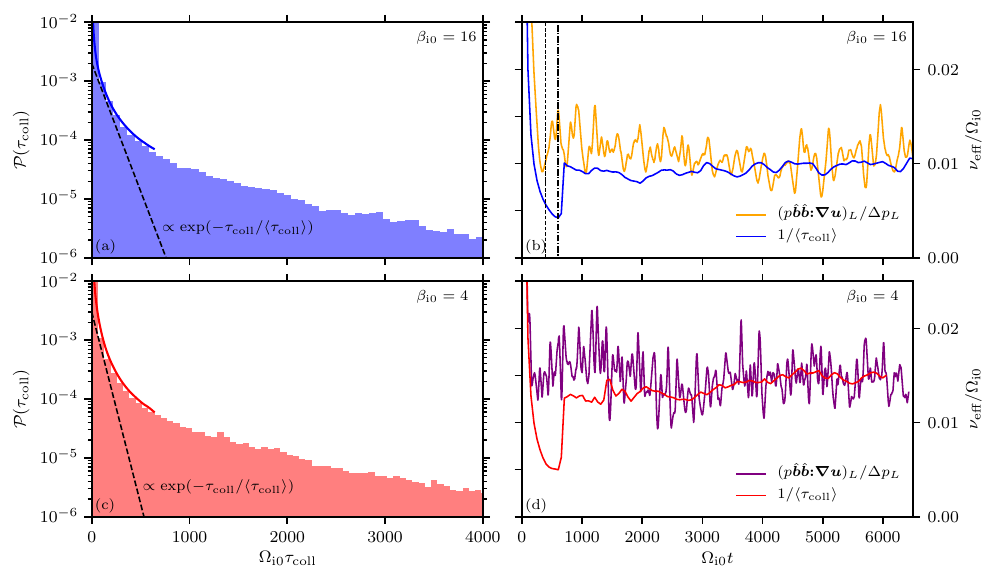}
    \caption{Histograms of collision times of tracked particles in the $\beta_{\rm i0} = 16$ (a) and $\beta_{\rm i0} = 4$ (c) simulations averaged over the duration of the simulations. Solid lines denote the probabilities obtained with the collisionality diagnostic described in~\S\ref{sec:collisionality}, which uses all particles in the simulation (\S\ref{sec:collisionality_2}). (b,d) Comparison of effective collisionalities obtained by computing $\Delta p_{\bs{k}}$ and $(p \ROS)_{\bs{k}}$ at large scales ($k_\perp \ell_{\nu,\rm eff} < 1$, see Figure~\ref{fig:phase_shift}) and those obtained from $\langle\tau_{\rm coll}\rangle$ based on the distributions in panels (a) and (c). A good agreement between these independent methods indicates that the collisionality is similar to the Braginskii estimate~\eqref{eq:nu_braginskii}. Vertical dashed and dot-dashed lines in panel (b) show the values of time for the snapshots in Figures~\ref{fig:slice_mirror} and \ref{fig:slice_IC}.}
    \label{fig:collisionality}
\end{figure*}

Despite their being comparable in magnitude, $p \ROS$ and $\nu_{\rm eff} \Delta p$ are not directly proportional to one another, as in the standard Braginskii closure. Instead, there is a phase difference between the two, a feature that may be explained as follows. If the local rate of strain behaves as $S(t) = S_0 \exp(- {\rm i} \omega t)$, and $\Delta(t=0) = 0$, then equation~\eqref{eqn:cgl} has the simple solution
\begin{equation}
    \Delta(t) = \frac{3 S_0}{-{\rm i}\omega + 3\nu} \, {\rm e}^{-{\rm i} \omega t} = \frac{S(t)}{\sqrt{\nu^2 + \omega^2/9}} \, {\rm e}^{{\rm i} \phi},\label{eqn:phase_lag}
\end{equation}
where $\cos \phi = \nu/\sqrt{\nu^2 + \omega^2/9}$ and $\phi$ is the phase lag between the pressure anisotropy and the rate of strain. This phase lag is relatively small at the injection scale, where $\nu \sim \beta \omega \gg \omega$, but becomes increasingly larger towards the smaller scales. Non-zero phase lag makes the energy transfer due to effective viscosity non-sign-definite, and makes the behavior of kinetic plasma different from a Braginskii-MHD plasma.

Another effect neglected in the Braginskii model is the growth of  small-scale magnetic fields associated with kinetic micro-instabilities (mostly firehose). Kinetic instabilities increase the ion-Larmor-scale contribution to the shear $\hat{\bb{b}}\hat{\bb{b}}\,\bb{:}\,\grad\bb{u}$ and cause it to deviate from $\Delta p$. Structure-function analysis of the rates of strain (not shown) suggests that $\hat{\bb{b}}\hat{\bb{b}}\,\bb{:}\,\grad\bb{u}$ peaks at $k_\parallel \rho_{\rm i0} \approx k_\perp \rho_{\rm i0} \approx 0.4$, which corresponds to the wavenumber of fastest growth for the oblique firehose instability~\citep{Kunz2014}.

\subsubsection{Effective collisionality from particle motion}
\label{sec:collisionality_2}

Figure~\ref{fig:phase_shift} provides just one way of estimating the effective collisionality of an (otherwise collisionless) high-$\beta$ plasma. This method relies on the assumption that pressure anisotropy evolves according to equation~\eqref{eqn:cgl}, which is not necessarily true if strong heat fluxes are present. To provide an independent measurement of the effective collisionality, we examine to what extent the magnetic moments $\mu$ of the particles are conserved. In the absence of scattering, the only way to change $\mu$ is through non-adiabatic heating or cooling. Although there has not yet been a self-consistent study of heating in collisionless high-$\beta$ plasmas, existing studies suggest that non-adiabatic heating is small; e.g., stochastic heating is suppressed at $\beta \gg 1$~\citep{Hoppock2018,Cerri2021}. Additionally, in the absence of pressure anisotropy, Landau damping can dissipate a significant portion of the cascade~\citep{Kawazura2018}, leaving little energy for additional non-adiabatic channels. 

An example of this process is shown in Figure~\ref{fig:track}. The black lines in panels (a)--(d) show the trajectory of a particle from the $\beta_{\rm i0} = 16$ simulation (see~\S\ref{sec:parameters}) in the planes perpendicular and parallel to the guide field, plotted over a snapshot of fluctuations in magnetic-field strength (a,b) and in pressure anisotropy (c,d). The fluctuations themselves evolve over time, while panels (a)--(d) only show them at a fixed moment; this should not present a problem of interpretation given that, at high plasma $\beta$, the fluctuations evolve much more slowly than the particles stream across them ($k_\parallel v_{\rm A} \ll k_\parallel v_{\rm th}$). This particular particle has been chosen because it moves through a region with firehose-unstable pressure anisotropy, and the magnetic-field slices exhibit clear firehose fluctuations. These fluctuations cause the particle's magnetic moment to break by pitch-angle scattering (panel~e), during which the total energy of the particle is almost constant (panel~f).

\begin{figure*}
    \centering
    \includegraphics[width=\textwidth]{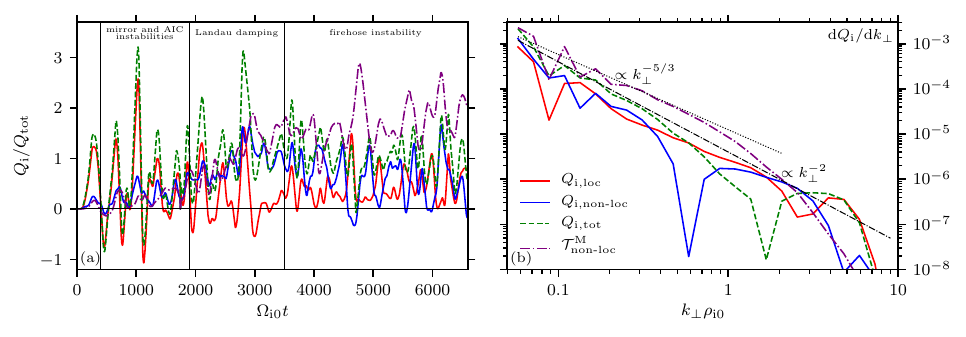}
    \caption{The dependence of anisotropic viscous heating, $\bb{\Pi}\,\bb{:}\,\grad \bb{u}$, in the $\beta_{\rm i0} = 16$ simulation on time (a) and wavenumber (b). Heating is separated into local (red) and non-local (blue) components, as defined in~\S\ref{sec:heating}, with total heating indicated by the green dashed line. The purple dot-dashed line denotes the non-local transfer from kinetic to magnetic energy, which we attribute to the growth of small-scale firehose fluctuations.}
    \label{fig:heat_vs_k}
\end{figure*}

One way of measuring collisionality from particle trajectories is to look at the particles' magnetic moments as functions of time, and compute a histogram of times $\tau_{\rm coll}$ needed for the magnetic moment of each tracked particle to change by one factor of e. In Figure~\ref{fig:collisionality}(a,c), we show such a histogram, computed using the trajectories of ${>}{10^4}$ tracked particles from our $\beta_{\rm i0} = 16$ and $\beta_{\rm i0} = 4$ simulations. Solid lines in this figure show the distribution functions of $\tau_{\rm coll}$, but evaluated during the simulation time using all available particles (${>}{10^{11}}$ particles in total)~\footnote{Unfortunately, the $\beta_{\rm i0}=16$ simulation with double-mode driving did not have tracked particles, and the $\beta_{\rm i0}=16$ simulation with single-mode driving did not have the collisionality diagnostic. Therefore, in this Figure, we combine diagnostics from these two different simulations. The simulations have similar distributions of $\tau_{\rm coll}$.}. These solid lines are then used to evaluate $\langle \tau_{\rm coll} \rangle$ at different times during the simulations. The effective collisionality is then obtained as $\nu_{\rm eff} = 1/\langle \tau_{\rm coll} \rangle$. We use $1/\langle \tau_{\rm coll} \rangle$ as a definition of the effective collisionality because it is the maximum likelihood estimator for an exponential distribution (which is our hypothesis for collision events). Panels (b) and (d) in Figure~\ref{fig:collisionality} compare the effective collisionalities computed in this manner with those computed from the ratio of rate of strain and pressure anisotropy at large scales $L \gtrsim \ell_{\nu,{\rm eff}}$ (\S\ref{sec:collisionality_1}).

In the beginning of each simulation, the collisionality is smaller than the Braginskii estimate, which is to be expected because the simulation is not yet in quasi-steady state. After the micro-instabilities are triggered (vertical dashed and dot-dashed lines in Figure~\ref{fig:collisionality}b), the collisionality  grows rapidly, and its value in quasi-steady state is consistent with the Braginskii estimate. The values of collisionality are also consistent with our expectations from~\S\ref{sec:theory}. Namely, for $\beta_{\rm i0} = 16$,
\begin{equation}
    \nu_{\rm eff}  \sim \beta_{\rm i0} \hat{\bb{b}}\hat{\bb{b}}\,\bb{:}\,\grad\bb{u}|_{\rm max} \sim \beta_{\rm i0} {\rm M}_{\rm A}^{3}\frac{v_{\rm A0}}{L_\perp} \approx 0.01 \Omega_{\rm i0}. \label{eq:nu_braginskii}
\end{equation}
Although the collisionality is expected to scale proportionally to $\sqrt{\beta_{\rm i0}}$ [based on equation~\eqref{eq:nu_braginskii} with $L_\perp \propto \rho_{\rm i0}$], we found a larger value of collisionality in the $\beta_{\rm i0}=4$ simulation than in the  $\beta_{\rm i0} = 16$ simulation. We attribute this to differences between Mach numbers and the quasi-steady-state pressure anisotropies in the simulations (the $\beta_{\rm i0} = 4$ simulation ends up closer to the firehose threshold, see Figure~\ref{fig:anisotropy}). If positive pressure anisotropies are mediated by the AIC instability with threshold ${\propto}1/\sqrt{\beta}$, then $\nu_{\rm eff} = \sqrt{\beta_{\rm i0}} \hat{\bb{b}}\hat{\bb{b}}\,\bb{:}\,\grad\bb{u}|_{\rm max}$, and thus both runs should have the same collisionality. However, our simulations are continuously driven, and after a long enough time, the firehose instability is triggered, which increases the collisionality to $\beta_{\rm i0} \hat{\bb{b}}\hat{\bb{b}}\,\bb{:}\,\grad\bb{u}|_{\rm max}$.

\begin{figure}
    \centering
    \includegraphics[width=\columnwidth]{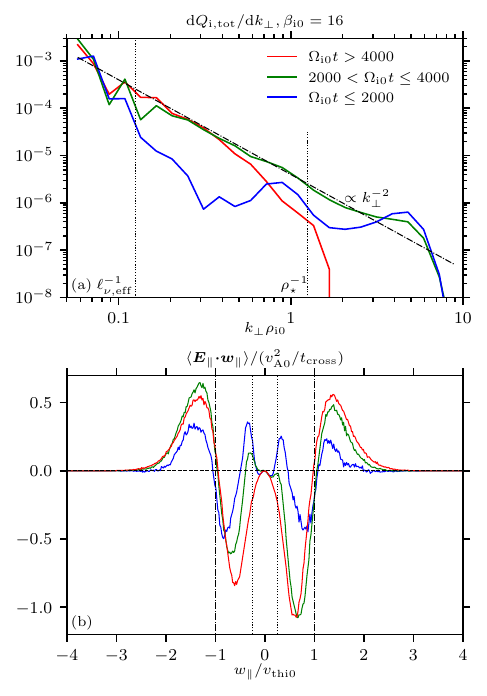}
    \caption{(a) Wavenumber dependence of particle energization during three time intervals in the $\beta_{\rm i0} = 16$ simulation. The first two stages have a considerable amount of heating at $k_\perp \rho_\star\sim 1$, where Landau damping is expected to be important. Sub-viscous heating is suppressed in the quasi-steady state, and most of the heating occurs close to the effective viscous scale. (b) Phase-space dependence of parallel energization of particles during the same time intervals. The peaks at $w_\parallel \sim \pm v_{\rm A0}$ (vertical dotted lines), indicative of Landau damping, are present during the early stages, but disappear in the quasi-steady state.}
    \label{fig:heat_vs_k_vs_t}
\end{figure}

\subsection{Ion heating}\label{sec:heating}

In~\S\ref{sec:theory}, we presented theoretical arguments, subsequently supported by our numerical results presented in~\S\S\ref{sec:viscosity} and \ref{sec:collisionality}, that suggested that the viscous scale in collisionless high-$\beta$ turbulence was close to the outer scale. Here we ask whether this means that most of the heating happens at the effective viscous scale, rather than at kinetic scales. Heating close to kinetic scales has been found numerically in low-$\beta$ kinetic turbulence~\citep{Arzamasskiy2019,Cerri2021,Squire2022}, and also in gyro-kinetic simulations of high-$\beta$ turbulence~\citep{Kawazura2018}. In the latter, it came from the Landau damping of Alfv\'en waves, expected to peak at a scale~\citep{Howes2006,Kunz2018,Kawazura2018}
\begin{equation}\label{eq:rho_star}
    \rho_{\star}\equiv (3/4\pi^{1/4}\sqrt{2})\beta_{\rm i}^{1/4}\rho_{\rm i}.
\end{equation}
Both low-$\beta$ simulations and gyrokinetic studies lack dynamically important viscous stresses, which can cause a significant portion of the cascade to be dissipated at the viscous scale. The precise amount of such dissipation might be difficult to estimate given the dynamical back-reaction of the parallel viscous stress, which tends to re-arrange fields so as to reduce the amount of parallel viscous heating~\citep{Squire2019,Kempski2019}. In this Section, we ask {\it whether most of the heating happens at small scales due to Landau damping or at large scales due to the pressure-anisotropic viscous stress, and show that the latter is the case in our simulations}.

We first note that the velocity spectrum in Figure~\ref{fig:spectrum_steady_state} has a slope steeper than $-5/3$, instead closer to $-2$. There are two potential explanations: either the non-linear interactions of turbulent eddies are modified in such a way as to steepen the spectrum, or the large-scale ion heating causes dissipation of a considerable portion of the cascade energy flux. As we explained in \S\ref{sec:viscosity}, for the former explanation to be valid, the conservative critically balanced cascade should satisfy $k_\parallel \delta u_k^2 \sim {\rm const}$, which for a $k_\perp^{-2}$ spectrum means that $k_\parallel \propto k_\perp$. As we showed in Figure~\ref{fig:slice_steady_state}(e,f), this is not the case, as $k_\parallel$ scales as $k_\perp^{1/2}$ in the  quasi-steady state. This means that the cascade is not conservative, and some part of the cascade is dissipated as ion heating. Our energy transfer function for the viscous stress (Figure~\ref{fig:tf_2d}) also suggests considerable dissipation in the inertial range.

To determine the wavenumber dependence of the ion heating $Q_{\rm i}$, we employ the energy-transfer functions for the thermal-energy equation (see \S\ref{sec:energy_transfer} and Appendix~\ref{sec:appendix_tf}), and separate the total heating into its local ($q_\perp = k_\perp$) and non-local ($q_\perp \ne k_\perp$) components averaged over the quasi-steady state. These averages are computed from two-dimensional transfer functions by summing over $q_\perp$-shells, $\mathcal{T}_{\textrm{non-loc}} \equiv \sum_{q_\perp \ne k_\perp} \mathcal{T}_{q_\perp\rightarrow k_\perp}$ and $\mathcal{T}_{\rm loc} \equiv \mathcal{T}_{k_\perp\rightarrow k_\perp}$. Such a definition can, in principle, depend upon the choice of wavenumber shells; for simplicity, we ignore this dependence, while noting that the local and non-local contributions to the transfer functions seen in Figure~\ref{fig:tf_2d} are quite distinct. Figure~\ref{fig:heat_vs_k} shows the dependence of local and non-local ion heating $\bb{\Pi}\,\bb{:}\,\grad \bb{u}$ on time (a) and on perpendicular wavenumber (b) for the $\beta_{\rm i0} = 16$ simulation. The simulation starts with most of $\bb{\Pi}\,\bb{:}\,\grad \bb{u}$ being local. This is the reversible energy transfer between bulk-kinetic and thermal energies; indeed, it changes sign several times during this stage of the simulation. After sufficiently negative pressure anisotropy has built up, so that the simulation becomes unstable to the firehose instability, the non-local ion heating becomes the dominant component of the energy transfer. During this stage, non-local heating approximately follows the non-local energy transfer between kinetic and magnetic energies due to firehose instability. In the quasi-steady state, there are comparable amounts of local and non-local heating, while the effective collisionality makes the transfer of energy due to anisotropic viscosity, $\bb{\Pi}\,\bb{:}\,\grad \bb{u}$, irreversible. The quasi-steady-state value of $\mathcal{T}^{\rm M}_{\textrm{non-loc}}$ is comparable to, or even larger than, the total heating, which indicates that non-local transfer due to firehose growth is comparable to the overall energy flux in the system and that the firehose instability is important in the quasi-steady state. We interpret all non-local transfer due to magnetic tension ($\mathcal{T}^{\rm M}_{\textrm{non-loc}}$) as firehose growth (see \S\ref{sec:energy_transfer}).

The steep spectrum of ${\rm d}Q_{\rm i}/{\rm d}k_\perp$ implies that most of the energy is dissipated at large scales, close to the effective viscous scale (and hence outer scale) of the system. Such dissipation is not present in low-$\beta$ simulations of collisionless Alfv\'{e}nic turbulence~\citep{Arzamasskiy2019,Cerri2021}, in which heating typically peaks at sub-ion scales. That being said, there is some dissipation at small scales as well: the spectrum of ${\rm d}Q_{\rm i}/{\rm d}k_\perp$ continues even after the cutoff of the kinetic-energy spectrum. There are several heating mechanisms that can operate in this range. Cyclotron heating, which can be important at $\beta \sim 1$~\citep{Arzamasskiy2019}, is expected to have a localized peak at the wavenumber at which the frequency of kinetic Alfv\'en waves $\omega_{\rm KAW} \sim \Omega_{\rm i}$; this behavior is inconsistent with the results in Figure~\ref{fig:heat_vs_k} (assuming Alfv\'enic nature of sub-$\rho_{\rm i}$ fluctuations, which is not exactly true if firehose instability is present). Similarly, stochastic heating~\citep{Chandran2010,Hoppock2018,Cerri2021} is expected to have a localized peak at $k_\perp \rho_{\rm i0} \sim 1$. Both of these mechanisms are expected to be relatively unimportant at $\beta \gg 1$. It is likely that the sub-viscous dissipation in our high-$\beta$ simulations is caused instead by Landau damping, which is the dominant energization mechanism seen in gyrokinetic simulations at high $\beta$~\citep{Kawazura2018}. That being said, the importance of Landau damping changes as the simulation progresses. Figure~\ref{fig:heat_vs_k_vs_t} shows the wavenumber and velocity dependence of ion energization during three time intervals: the mirror and AIC stages, $\Omega_{\rm i0 }t \leq 2000$; the Landau-damping stage, $2000<\Omega_{\rm i0 }t \leq4000$; and the quasi-steady state, $\Omega_{\rm i0 }t > 4000$. The first two intervals have considerable heating near $k_\perp \rho_\star \sim 1$ [recall the definition of $\rho_\star$ given by equation~\eqref{eq:rho_star}]. In contrast, Larmor-scale heating is considerably suppressed in the final stage \footnote{All intervals have significant energization at the outer scale, which is likely due to adiabatic processes such as the overall growth of the magnetic-field strength in the box.}. The phase-space energization is defined as the product $\langle \bb{E}_\parallel \bcdot \bb{w}_\parallel\rangle$ of the local parallel electric field and the parallel component of particle peculiar velocity ($\bb{w}\equiv \bb{v} - \bb{u}$) averaged over all particles within a given region of the velocity space (Figure~\ref{fig:heat_vs_k_vs_t} only shows the dependence on $w_\parallel$). The origin of the sign reversal around $w_{\parallel} \sim \pm v_{\rm thi0}$ remains uncertain, but could relate to the Landau-damping of acoustic fluctuations or could be a signature of the viscous heating (though if so, exactly how this arises remains poorly understood). The Landau-damping signatures at $w_\parallel \sim \pm v_{\rm A0}$ are present at early times, but recede as the simulation progresses. This indicates that {\it Landau damping is suppressed (but still important), and a significant fraction of the heating in the quasi-steady state is due to the pressure-anisotropic viscous stress}. We measure large-scale viscous heating to be responsible for approximately one-half of the overall heating rate.

We defer an investigation of how heating and scattering affect the ion distribution function to Appendix~\ref{sec:appendix_dist}.

\section{Discussion}\label{sec:discussion}

\subsection{Simulation dynamics and effective viscosity}

In this paper, we explored the evolution of turbulent fluctuations in a collisionless, high-$\beta$ plasma. The initially driven fluctuations become unstable to mirror and, later, to AIC instabilities. These instabilities result in ion-Larmor-scale perturbations of the magnetic field, which cause ions to scatter with a characteristic scattering frequency ${\sim}S \beta$, where $S \equiv \hat{\bb{b}}\hat{\bb{b}}\,\bb{:}\,\grad \bb{u}$ is the growth rate of the magnetic-field strength. This scattering limits the (otherwise adiabatically driven) pressure anisotropy. The average value of the pressure anisotropy ultimately becomes negative (i.e., $p_\parallel > p_\perp$), thereby triggering the firehose instability. The pressure anisotropy then fluctuates across the box and has a broadband spectrum that peaks at the viscous scale, which, for the effective collisionality $\nu\sim S\beta$ and Alfv\'{e}nic Mach number ${\rm M}_{\rm A}\sim 1$, is always comparable to the driving scale. The root-mean-square value of pressure anisotropy is consistent with the Braginskii value of $\nu^{-1} \hat{\bb{b}} \hat{\bb{b}}\,\bb{:}\,\grad \bb{u}$ with $\nu$ measured directly from the particles, although there is a phase lag between it and the parallel rate of strain, which makes the dynamical impact of pressure anisotropy somewhat different than found in Braginskii-MHD \citep{Squire2019}. The parallel viscous stress associated with the instability-regulated pressure anisotropy leads to irreversible dissipation, which peaks at a scale $\ell_{\nu,\rm eff}$ that is close to the outer scale of the turbulence, and steepens the kinetic-energy spectrum below that scale to a spectral slope close to $-2$.

\subsection{Ion energization}

At high values of $\beta$, the viscous heating, to which we attribute the steepening of the kinetic-energy spectrum, dissipates the majority of the cascade energy flux. This heating mechanism should persist, no matter the scale separation in the system, as long as the outer-scale fluctuations are above the Alfv\'en-wave interruption limit (i.e., as long as their amplitudes are such that the pressure anisotropy that they adiabatically produce ventures beyond the $\beta$-dependent kinetic-instability thresholds). In addition to large-scale viscous heating, there is considerable heating at sub-viscous scales, especially during the early stages of the simulations. We attribute this heating, which peaks at $k_\perp \rho_{\rm i} \sim \beta^{-1/4}$ [see equation \eqref{eq:rho_star}], to Landau damping, the dominant mechanism in gyro-kinetic simulations \citep{Kawazura2018}. The ion-to-electron heating ratio is ${\sim}5$--$10$ in all our simulations (electron heating is measured via the hyper-resistive dissipation at $k_\perp \rho_{\rm i0} \gg 1$). There is also considerable non-local energy transfer from driving scales to kinetic scales due to the firehose instability (comparable to overall energy flux; see Figure~\ref{fig:heat_vs_k}).

\begin{figure*}
    \centering
    \includegraphics[width=\textwidth]{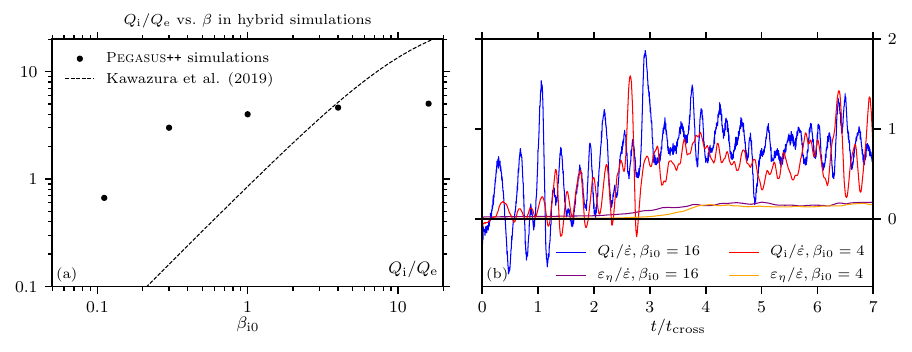}
    \caption{(a) The ratio of ion to electron energization, $Q_{\rm i}/Q_{\rm e}$, as a function of plasma $\beta$ from several hybrid-kinetic simulations ($\beta_{\rm i0} = 1/9$ simulation from \citep{Cerri2021}; $\beta_{\rm i0} = 1$ and $0.3$ simulations from \citep{Arzamasskiy2019}; and $\beta_{\rm i0} = 16$ and $4$ simulations from this paper). Ion heating is measured directly from the evolution of the thermal energy of the particles. Electron heating is inferred from hyper-resistive dissipation near the grid scale. For comparison, the dashed line shows the results from the hybrid-gyrokinetic simulations of Ref.~\citep{Kawazura2018}. (b) Time dependence of ion heating ($Q_{\rm i}$) and hyper-resistive dissipation ($\varepsilon_{\eta}$) in the high-$\beta$ hybrid-kinetic simulations. All lines are normalized to the total energy dissipation $\dot\varepsilon$ averaged over the final $3.5t_{\rm cross}$ of each simulation.}
    \label{fig:Q_vs_beta}
\end{figure*}

For studies of high-$\beta$ turbulence in fully collisionless plasmas, there is an important physical ingredient missing from our simulations: realistic electron physics. Imagine a turbulent plasma composed of collisionless ions and electrons. The large-scale fluctuations make the distribution functions of both ions and electrons anisotropic, and at sufficiently large $\beta$ unstable to pressure-anisotropy-driven instabilities. If the quasi-steady-state anisotropy of both species is Braginskii-like, with $\Delta p/p \sim \nu^{-1} \hat{\bb{b}}\hat{\bb{b}}\,\bb{:}\,\grad \bb{u}$ -- as it is for ions in our simulations -- the ion-to-electron heating ratio should be
\begin{equation}\label{eq:qi_qe_discussion}
    \frac{Q_{\rm i}}{Q_{\rm e}} \sim \frac{p_{\rm i}/\nu_{\rm i}}{p_{\rm e}/\nu_{\rm e}}.
\end{equation}
Therefore, the partition of energy is determined by the effective collisionality of the species. If the collisionality of both species is $\nu \sim S \beta$, then $p/\nu \propto \beta/\nu = {\rm const}$ is independent of both $\beta$ and the species. It is therefore possible that ions and electron receive the same amounts of energy, which contradicts, e.g., the models used in the interpretation of EHT images of black-hole accretion flows in M87 and around Sgr A$^\star$~\citep{Ressler2015,Chael2018}.This estimate depends on the exact thresholds of the instabilities that regulate the particle velocity distributions. For example, \citet{Sharma2007} argue that, for AIC and electron-whistler instabilities, $Q_{\rm i}/Q_{\rm e} \sim 10$,  consistent with some theories of radiatively inefficient black-hole accretion flows.

Additionally, in the presence of cooling, the electron temperature can decrease due to radiation. This would lead to a decrease in $T_{\rm e}$, and thus $\beta_{\rm e}$. If $\beta_{\rm e}$ decreases to a point where the plasma is stable to electron micro-instabilities, this can lead to a state with $Q_{\rm i}/Q_{\rm e} \gg 1$ and thus $T_{\rm i}/T_{\rm e} \gg 1$. The resulting large temperature ratios may persist, as there are no known collisionless mechanisms for efficient electron-ion thermal coupling \citep{Zhdankin2021}.

One particularly important application of the results of this section is to the interpretation of EHT images of black-hole accretion flows. This interpretation involves carrying out general-relativistic magnetohydrodynamic simulations with some prescription for the heating rate of the electrons \citep{Ressler2015,Chael2018}, which is typically informed from gyrokinetic calculations \citep{Howes2008heating,Kawazura2018}. In Figure~\ref{fig:Q_vs_beta}, we display a summary plot of the ion-to-electron heating ratio obtained from our hybrid-kinetic simulations done with \textsc{Pegasus}\texttt{++} as a function of plasma $\beta$. The low-$\beta$ values are taken from our earlier work on ion energization in strong Alfv\'{e}nic turbulence~\citep{Arzamasskiy2019,Cerri2021}, while the high-$\beta$ points represent the results of this paper. Averaged over the final $3.5 t_{\rm cross}$, about $83\%$ of the total cascaded energy is absorbed by ions in the $\beta_{\rm i0} = 16$ simulation, and about $82\%$ at $\beta_{\rm i0} = 4$. The remaining energy flux cascades further until it is removed by hyper-resistivity near the grid scale. We estimate the electron heating in hybrid kinetics as this hyper-resistive dissipation, $\varepsilon_\eta$. We also plot (dashed line) the predicted $Q_{\rm i}/Q_{\rm e}$ from a series of simulations with gyrokinetic ions and fluid electrons~\citep{Kawazura2018}. Note that gyrokinetics assumes $\delta B/B \ll 1$, while the fluctuations in our simulations have finite amplitudes. Our results differ significantly from these hybrid-gyrokinetic predictions, which indicates the importance of including non-adiabatic processes (such as kinetic micro-instabilities and non-adiabatic heating channels) that are ordered out of gyrokinetics (the only available dissipation channels in gyrokinetics are parallel Landau damping and non-linear perpendicular phase mixing).

Finally, it is important to understand the caveats related to using Figure~\ref{fig:Q_vs_beta}a for black-hole accretion models. Other than missing electron physics, an important limitation of our work is limited scale separation between the energy-injection scale and the dissipation scale of the cascade. For the low-$\beta$ runs~\citep{Arzamasskiy2019,Cerri2021}, the effective scale separation is ${\sim}10^4$ (the separation between $\rho_{\rm i}$ and the scale at which $\delta u/v_{\rm A} \sim 1$). Such a scale separation is realistic for the solar wind, but not for accretion flows, where it is expected to be ${\sim}10^{10}$. Dissipation mechanisms used to explain the low-$\beta$ results (stochastic and cyclotron heating) are expected to diminish with scale separation, so it is unclear whether the conclusions from Ref.~\citep{Arzamasskiy2019,Cerri2021} will hold at scale separations relevant to black-hole accretion. That being said, astrophysical turbulence can be imbalanced (e.g., the solar-wind turbulence is measured to be imbalanced \citep{McManus2020}). Recent work on imbalanced cascades~\citep{Meyrand2021,Squire2022} concluded that the imbalanced portion of the energy flux could not cascade beyond the ion-Larmor scale and was eventually dissipated through AIC heating. Such dissipation is expected to be controlled solely by the degree of imbalance and not by the scale separation in the cascade. 

For high-$\beta$ turbulence, the results of this paper (\S\ref{sec:heating}) indicate that the majority of dissipation happens at the outer scale of the cascade. Therefore, ion heating may depend on the properties of the forcing, and may require a better understanding of realistic turbulence injection (e.g., through kinetic magnetorotational instability~\citep{Kunz2016,Kawazura2021,Bacchini2022}). The results from high-$\beta$ hybrid-kinetic simulations are likely to depend upon the amplitude of the forcing and on the scale separation (e.g., through their impact on spectral anisotropy at ion-Larmor scale). To test the latter, we have conducted a test for the amplitude dependence by running a $\beta_{\rm i0} = 16$ simulation with $\delta u_L/v_{\rm A}\sim 1$, having a lower ion-scale spectral anisotropy than other runs used in this work. In this simulation, a slightly larger fraction of the cascade rate, ${\sim}90\%$, is dissipated on ions.

\subsection{Dependence on scale separation}

Given the limited size of our simulations, it is important to understand whether our results are expected to hold at scale separations relevant to astrophysical systems. One important (although transient) feature of our runs is Landau damping, which eventually pushes the pressure anisotropy over the firehose-instability threshold by raising the parallel temperature. If the effective collisionality from the instabilities is strong enough to interfere with the maintenance of the Landau resonance (namely, $\nu_{\rm eff} \gg k_\parallel v_{\rm thi}$), then the Landau damping can be shut off. From our estimates in \S\ref{sec:theory}, $\nu_{\rm eff}\sim \beta {\rm M}_{\rm A}^3 v_{\rm A}/L$, and thus at $k_\parallel \rho_{\star}  \sim 1$, where Landau damping is expected to become important, $k_\parallel v_{\rm thi}/\nu_{\rm eff} \sim (L/\rho_{\rm i})/(\beta_{\rm i}^{3/4} {\rm M}_{\rm A}^3 ) \gg 1$, the inequality following from the typically enormous astrophysical scale separation between $L$ and $\rho_{\rm i}$. Therefore, Landau damping is expected to be important (arguably more important than in our simulations) and the plasma will approach the  firehose-instability threshold in approximately one large-scale dynamical time (assuming that the entire cascade with ${\rm M}_{\rm A}\sim 1$ is dissipated as parallel heating; our simulations approach it in ${\sim}4t_{\rm cross}$ given the smaller fluctuation amplitude of ${\rm M}_{\rm A}\sim 0.5$).

In order to estimate the amplitude of firehose fluctuations at the Larmor scale, we can leverage some results from recent \textsc{Pegasus\texttt{++}} expanding-box simulations \citep{Bott2021,Bott2022}, which have shown for $\Omega_{\rm i}/S \gtrsim 30 \beta_{\rm i}^{3/2}$ that the kinetic firehose threshold is ${\approx}-1.4/\beta_{\rm i}$ and that the instability saturates at an amplitude $(\delta B_\perp/B)_{\rm FH}^2 \sim (S/\Omega_{\rm i})^{1/2} \sim {\rm M}_{\rm A}^{3/2} (\rho_{\rm i}/L)^{1/2} \beta_{\rm i}^{-1/4}$ for $S \sim {\rm M}_{\rm A}^3 v_{\rm A}/L$. For comparison, the amplitude of fluctuations in an Alfv\'enic cascade with the Goldreich--Sridhar spectrum evaluated at $k_\perp \rho_{\rm i} \sim 1$ is $(\delta B_\perp/B)_{\rm AW}^2 \sim  {\rm M}_{\rm A}^2 (\rho_{\rm i}/L)^{2/3} \ll (\delta B_\perp/B)_{\rm FH}^2$ for ${\rm M}_{\rm A} \sim 1$ and $\rho_{\rm i}/L \ll \beta_{\rm i}^{-3/2}$. For a  $k_\perp^{-2}$ spectrum, $(\delta B_\perp/B)_{\rm AW}^2$ is even smaller at the ion-Larmor scale. We therefore conclude that in astrophysical systems, as found in our simulations, ion-Larmor-scale magnetic-field fluctuations are expected to be composed mostly of firehose modes rather than of Alfv\'enic fluctuations. Therefore, ion-Larmor-scale fluctuations in our simulations are similar to those expected in astrophysical systems.

\subsection{Observational implications}

Unfortunately, outside of the solar wind, observations of turbulence in high-$\beta$ systems are limited. And the solar-wind observations are in a slightly different regime than our simulations: the solar wind starts as a low-$\beta$ plasma, which expands and reaches the high-$\beta$ regime. In this situation, which we studied in Ref.~\citep{Bott2021}, the expansion is also an important source of pressure anisotropy. Nevertheless, the solar wind shows a pressure anisotropy with an average value close to zero, and with a spread consistent with the $1/\beta$ instability thresholds (although the collisional age at ${\sim}1$~au, being comparable to outer-scale dynamical timescales, is short enough to matter~\citep{Bale2009}). Our simulations show very similar behavior (see Figure~\ref{fig:anisotropy}), with a spread in the pressure anisotropy comparable to $1/\beta$ and an average value that is negative but smaller than the firehose threshold. 

Outside the solar wind, the most promising system in which high-$\beta$ turbulence can be studied is the ICM, which has $\beta\sim 100$. There have been several attempts to measure plasma-velocity fluctuations in the ICM. The X-ray observations presented in  Ref.~\citep{Zhuravleva2019} use Bremsstrahlung emission from the hot intracluster plasma to determine the density fluctuations, from which the velocity fluctuations are then inferred. Their energy spectrum is consistent with the Kolmogorov prediction and extends to scales considerably smaller than the viscous scale expected from Coulomb collisions alone. This implies that the effective collisionality of the ICM is appreciably enhanced. One explanation for this enhancement is scattering from kinetic micro-instabilities, which can be triggered for turbulence with Alfv\'enic Mach numbers ${\rm M}_{\rm A} \gtrsim 1/\beta \ll 1$. The value of collisionality required to explain the spectra in Ref.~\citep{Zhuravleva2019} is consistent with our analytical estimate (\ref{eqn:mfpeff}).

Our simulations show that, although the effective viscous scale is large, the spectrum in this case is closer to Kolmogorov than $\grad^2 \bb{u}$ dissipation would imply (it steepens to approximately $k_\perp^{-2}$ instead of exhibiting an exponential decrease; current observations \citep{Zhuravleva2019} cannot distinguish between $-5/3$ and $-2$ spectral slopes). Ref.~\citep{Li2020} used optical emission from cold gas in the ICM to measure the spectrum of plasma velocity. Their observations show a slightly steeper spectrum than $-5/3$; for some clusters, it is close to $-2$. Our simulations predict similar spectra: the effective viscosity does not produce an exponential cutoff in the spectrum, but rather steepens it slightly. Optical measurements are much more precise than those taken in the X-ray, but their connection to turbulence in the bulk ICM is unclear (optical measurements are dominated by the interiors of cluster cores, which are expected to be much more collisional than ICM outskirts). Future observations are required to determine better the relationship between our simulations and the ICM turbulence.

Of course, even though we have a testable prediction for the slope of the sub-viscous spectra, the scale separation used in our simulations is much smaller than in actual astrophysical systems; our simulations have only ${\sim}1$ decade in scale between $\ell_{\nu,\rm eff}$ and $\rho_{\rm i}$, which is much smaller than the ${\sim}12$ decades in the ICM. Relatively small scale separation leads to firehose fluctuations being produced relatively close to $\ell_{\nu,\rm eff}$, which may impact the spectral slopes and the comparison between the rate of strain and the pressure anisotropy. However, we are encouraged that our numerical results agree well with the analytical estimates presented in \S\ref{sec:theory}, which we expect to hold for any scale separation. Given the steep scaling of the computational cost of kinetic simulations with scale separation (as the fourth power), future efforts should concentrate on the development of realistic sub-grid models for kinetic physics in high-$\beta$ plasmas.

\section{Summary}
\label{sec:summary}

We have presented analytical estimates of effective collisionality and viscosity in collisionless high-$\beta$ turbulence and tested those estimates by performing first-principles hybrid-kinetic simulations. We explored the interplay between local non-linear turbulent cascades and kinetic instabilities, which tap the free energy of large-scale deviations from local thermodynamic equilibrium (e.g., pressure anisotropies) to produce small-scale magnetic-field fluctuations. Our results can be summarized as follows.
\begin{itemize}\itemsep -0.25ex
    \item Large-scale fluctuations (continuously driven by a random, incompressible force in our simulations) generate pressure anisotropy through approximate adiabatic invariance and ultimately excite rapidly growing kinetic instabilities, thereby transferring energy non-locally to small (ion-Larmor) scales.
    \item At the beginning of the simulations, the main kinetic instabilities are mirror and AIC, because the magnetic-field strength in our box is mostly increasing, which drives positive pressure anisotropy.
    \item Turbulence reaches an intermediate steady state, in which the cascade energy is dissipated primarily by Landau damping, causing parallel heating.
    \item The parallel heating pushes the system towards the firehose instability threshold.
    \item In the quasi-steady state, the turbulence is primarily mediated by the firehose instability with non-local energy transfer due to the growth of small-scale firehose fluctuations that is comparable to the overall energy flux. The mean pressure anisotropy is slightly negative, with some parts of the box below the kinetic-firehose threshold.
    \item Firehose instability creates small-scale magnetic fluctuations, which scatter particles. The effective collisionality in the quasi-steady state is consistent with an estimate based on incompressible Braginskii MHD with the pressure anisotropy regulated by firehose instability, {\em viz.,}~$\nu_{\rm eff}\sim \beta \ROS$.
    \item The effective viscous scale due to the firehose-induced collisionality is close to the driving scale of the turbulence. For a cascade with Alfv\'enic Mach number ${\rm M}_{\rm A}$, our analytical estimates suggest that the ratio of effective viscous scale and the outer scale is $\ell_{\nu}/L \sim {\rm M}_{\rm A}^{-3}$. In our simulations, ${\rm M}_{\rm A} \sim 1$ at the outer scale.
    \item In addition to small-scale dissipation due to Landau damping, there is a considerable amount of viscous heating at the effective viscous scale.
    \item Ion heating removes the majority of the cascading energy, ${\sim}$80--90\% for $\beta = 4$ and $16$.
    \item Viscous heating steepens the kinetic-energy spectrum of the turbulence. The spectrum is approximately $k_\perp^{-2}$. Large-scale viscous heating dissipates ${\approx}40{-}45\%$ of the cascade as ion heating.
    \item The magnetic-energy spectrum is shallower than $k_\perp^{-5/3}$ near the ion-Larmor scale due to the presence of firehose fluctuations.
\end{itemize}

This work provides the first self-consistent estimate of the effective viscosity in a turbulent collisionless high-$\beta$ plasma. It makes observational predictions, which can be tested in the solar wind~\footnote{An important source of pressure anisotropy in the solar wind is its expansion and the consequent dilution of the interplanetary magnetic field~\cite{LeerAxford1972}. See Ref.~\cite{Bott2021} for a recent discussion of how such expansion affects Alfv\'enic turbulence in a collisionless, magnetized plasma.} and in the intracluster medium, and raises important complications for models of radiatively inefficient black-hole accretion flows.


\begin{acknowledgments}

This work benefited from useful conversations with Christopher Chen, Jeremy Goodman, Kristopher Klein, Yuan Li, Christopher Reynolds, Anatoly Spitkovsky, James Stone, Vladimir Zhdankin, Ellen Zweibel, and especially Silvio Sergio Cerri and Archie Bott. Assistance from Daniel Gro{\v s}elj with computing structure functions is gratefully acknowledged. L.A.~was supported by the National Aeronautics and Space Administration (NASA) under Grant No.~NNX17AK63G issued through the Astrophysical Theory Program, a Harold W.~Dodds Honorific Fellowship from Princeton University, and the Institute for Advanced Study. Support for M.W.K.~was provided by NASA grant NNX17AK63G and Department of Energy (DOE) award DE-SC0019046 through the NSF/DOE Partnership in Basic Plasma Science and Engineering. Support for J.S.~was provided by Rutherford Discovery Fellowship RDF-U001804 and Marsden Fund grant UOO1727, which are managed through the Royal Society Te Ap\=arangi. E.Q.~was supported in part by a Simons Investigator award from the Simons Foundation. The work of A.A.S.~was supported in part by the UK EPSRC grant EP/R034737/1 and the UK STFC grant ST/W000903/1. High-performance computing resources were provided by: the Texas Advanced Computer Center at The University of Texas at Austin under allocation numbers TG-AST160068 and AST130058; and the PICSciE-OIT TIGRESS High Performance Computing Center and Visualization Laboratory at Princeton University. This work used the Extreme Science and Engineering Discovery Environment (XSEDE), which is supported by NSF grant number ACI-1548562.

\end{acknowledgments}

\bibliography{AKSQS22}

\appendix

\section{Mirror and AIC instabilities in $\bb{\beta_{\rm i0} = 16}$ simulation}
\label{sec:appendix_instabilities}

\begin{figure*}
    \centering
    \includegraphics[width=\textwidth]{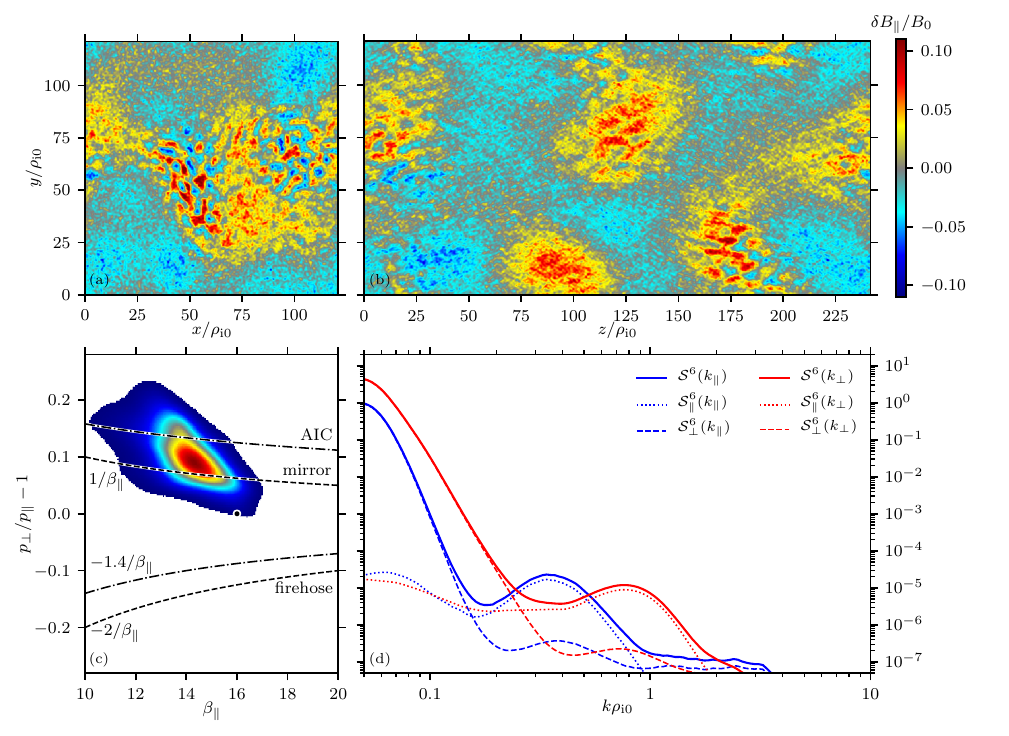}
    \caption{(a,b) Snapshots of parallel magnetic-field fluctuations $\delta B_\parallel$ (relative to $\bb{B}_0$) during the early stage of the $\beta_{\rm i0} = 16$ simulation ($t\approx 0.4 t_{\rm cross}$). Panel (a) shows a slice perpendicular to the guide field; panel (b) shows a slice along the guide field. There is a clear ``cross'' pattern in both slices, indicating oblique mirror fluctuations. (c) The distribution function of points in the box as a function of their $\beta$ and pressure anisotropy. Dashed lines show the thresholds for mirror \citep{ShapiroShevchenko1964,Barnes1966,SouthwoodKivelson1993} and fluid-firehose \citep{Rosenbluth1956,Parker1958b,Chandrasekhar1958,VedenovSagdeev1958} instabilities; dot-dashed line represents a threshold for the AIC instability \citep{SagdeevShafranov1960,Gary1994} and kinetic firehose instabilities \citep{Bott2021,Bott2022}. In this snapshot, the majority of the box is above the mirror threshold. The black dot indicates the initial position of the simulation box. (d) 6th-order structure function of the fluid velocity- and magnetic-field fluctuations. High-order structure functions are chosen to highlight intense small-scale fluctuations, which for this snapshot are mirror modes: oblique modes producing kinetic-range peaks in both parallel and perpendicular structure functions of parallel magnetic-field fluctuations of parallel magnetic field.}
    \label{fig:slice_mirror}
\end{figure*}

In the early stages of our simulations, the external driving excites large-scale modes, which produce large coherent patches of positive pressure anisotropy, owing to the conservation of the particles' magnetic moments. Once this pressure anisotropy grows above the ${\sim}1/\beta$ threshold, the mirror instability is triggered. This instability produces ``cross-patterned'' oblique structures in $\delta B_\parallel$, which act to reduce the magnetic-field strength locally. Particles trapped in these structures ``see'' an almost constant-in-time magnetic field, which prevents the pressure anisotropy of the trapped-particle population to grow beyond the mirror-instability threshold. Eventually the mirrors become large enough and their edges sharp enough to scatter particles~\cite{Kunz2014,Melville2016}.

\begin{figure*}
    \centering
    \includegraphics[width=\textwidth]{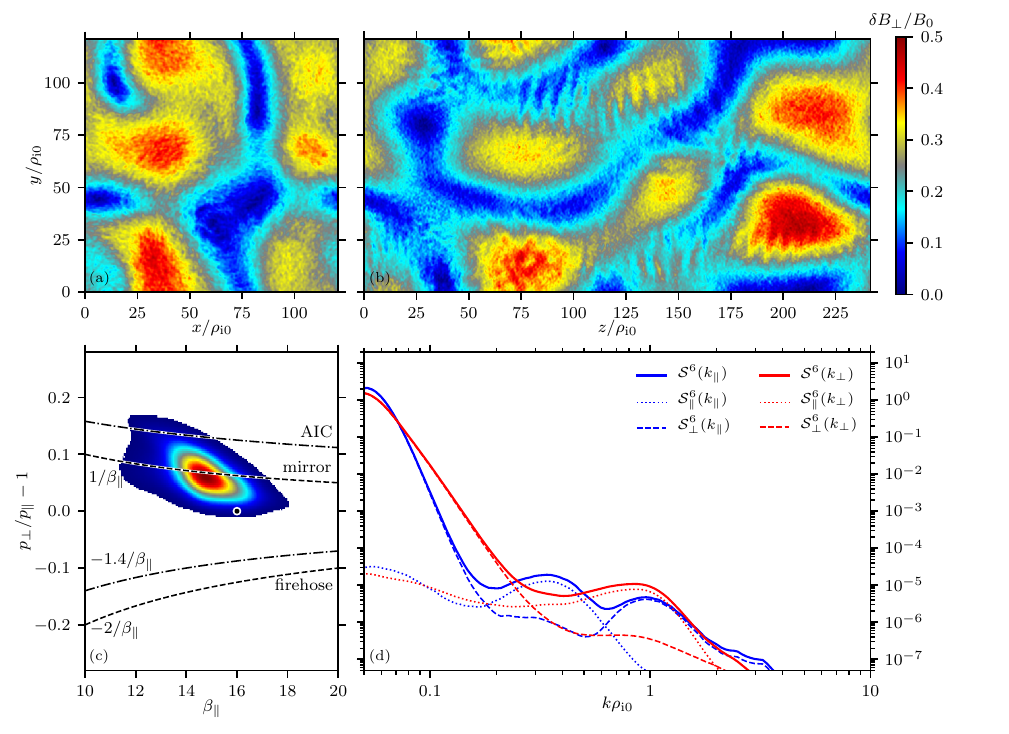}
    \caption{The same as Figure \ref{fig:slice_mirror}, but for a slightly later time $t\approx 0.6 t_{\rm cross}$, at which quasi-parallel fluctuations are manifest. The perpendicular component of magnetic field is shown instead of the parallel one. Those fluctuations are produced by the AIC instability. The structure functions show two predominant modes of fluctuations: oblique fluctuations in $\delta B_\parallel$ caused by mirror instability, and quasi-parallel fluctuations in $\delta B_\perp$, which are AIC waves.}
    \label{fig:slice_IC}
\end{figure*}

We illustrate this process in Figure~\ref{fig:slice_mirror}, which shows snapshots of $\delta B_\parallel$ in the planes perpendicular (panel~a) and parallel (panel~b) to the background magnetic field, two-dimensional histograms (panel~c) of pressure anisotropy and plasma $\beta$, and structure functions of the magnetic-field fluctuations (panel~d), defined as
\begin{equation}
    \mathcal{S}^n (\bb{\ell}) \equiv \langle |\bb{B}(\bb{x} + \bb{\ell})-\bb{B}(\bb{x}) |^n \rangle_{\boldsymbol{x}},\label{eqn:sf_tot}
\end{equation}
where $n$ is the order of the structure function and $\langle\,\cdot\,\rangle_{\boldsymbol{x}}$ represents the spatial average over the simulation domain. In addition to the full structure function (\ref{eqn:sf_tot}), we also compute the structure functions of magnetic-field fluctuations oriented parallel and perpendicular to the local, scale-dependent magnetic-field direction. To determine the latter, we define
\begin{equation}
    \bb{B}_{\rm loc}(\bb{\ell},\bb{x}) \equiv [\bb{B}(\bb{x} + \bb{\ell}) + \bb{B}(\bb{x})]/2,
\end{equation}
and then compute the structure functions of those fluctuating magnetic-field components parallel and perpendicular to $\hat{\bb{b}}_{\rm loc } \equiv \bb{B}_{\rm loc}/B_{\rm loc}$ \citep{Chen2011a}, e.g., $\mathcal{S}_\parallel$ represents the structure function computed using $\bb{B}_\parallel (\bb{\ell},\bb{x}) \equiv [\bb{B}(\bb{x})\bcdot\hat{\bb{b}}_{\rm loc}(\bb{\ell},\bb{x})]\hat{\bb{b}}_{\rm loc}(\bb{\ell},\bb{x})$, and $\mathcal{S}_\perp$ represents a structure function computed using $\bb{B}_\perp(\bb{\ell},\bb{x}) \equiv \bb{B}(\bb{x}) - \bb{B}_\parallel(\bb{\ell},\bb{x})$. For additional details concerning structure-function analyses of turbulence simulations, we refer the reader to Ref.~\cite{Cerri2019}.

Figure~\ref{fig:slice_mirror} shows a snapshot from the $\beta_{\rm i0} = 16$ simulation at a relatively early time. One can see several large-scale modes, which have not yet had time to shear one another. These large-scale fluctuations produce appreciable positive pressure anisotropy ($\Delta p > 0$), with a considerable fraction of the box being above both the mirror and AIC instability thresholds. We have chosen this particular snapshot because it highlights the mirror instability being triggered by pressure anisotropy -- throughout the box, one can see oblique fluctuations predominantly of $\delta B \approx \delta B_\parallel$. To examine these fluctuations further, we plot the sixth-order structure functions of $\delta B_\perp$ (blue) and $\delta B_\parallel$ (red). The high order of these structure functions is chosen to highlight localized, high-amplitude structures, such as those expected to be produced by kinetic micro-instabilities \citep{Kunz2014}. This analysis clearly shows that there is considerable magnetic energy at the driving scales of the simulation and some energy at the kinetic scales (due to instabilities), with very little energy in between. This is an indication that the energy stored in the anisotropic distribution function has been transferred non-locally from the driving scales (which have the largest pressure anisotropy) to the kinetic scales.

At the scale separations achieved in our simulations, trapping of particles by mirror fluctuations and the consequent regulation of the pressure anisotropy towards the mirror thresholds occur relatively slowly. Because of this, the unstable pressure anisotropy \citep{Kunz2014,Rincon2015,Melville2016} overshoots the mirror threshold enough to reach the AIC instability threshold (this is unlikely to occur in real astrophysical systems, which have significantly larger scale separations). Figure~\ref{fig:slice_IC} shows a simulation snapshot at $t\approx 0.6 t_{\rm cross}$, by which time the AIC instability has had enough time to grow, as indicated by the strong quasi-parallel fluctuations in $\delta B_\perp$ (cf.~Figure~\ref{fig:slice_mirror}). Examination of the structure functions indicates that this instability is cleanly separated from the previously triggered mirrors: in addition to ``bumps'' in the structure functions due to the mirror instability, an extra bump in $\delta B_\perp$ appears, with no obvious indication of interactions between the two instabilities. Finally, the histogram of pressure anisotropy and $\beta_\parallel$ shows that the pressure anisotropy has decreased from the previous snapshot even though the fluctuation amplitude at the driving scale has increased, suggesting that the instabilities have already back-reacted on the plasma and reduced its departures from isotropy. 

\section{Energy transfer in kinetic high-$\bb{\beta}$ turbulence}
\label{sec:appendix_tf}

In this Appendix, we summarize the energy-transfer analysis that we use to study non-local interaction in high-$\beta$ kinetic turbulence. Energy is injected by the external forcing in the form of bulk kinetic energy $\mathcal{E}_{\rm bulk} \equiv \varrho u^2/2$. Part of this energy is then converted into magnetic energy $\mathcal{E}_{\rm mag} \equiv B^2/8\pi$ through electromagnetic induction. This energy is ultimately dissipated by increasing the thermal energy of ions ($\mathcal{E}_{\rm th}\equiv p_\perp + p_\parallel/2$) or by hyper-resistivity at small scales.

The evolution of bulk kinetic energy follows from the momentum equation for ions (assuming gyrotropy of the ion distribution function):
\begin{align}
    \frac{\partial \varrho \bb{u}}{\partial t} = &-\grad \bcdot \left[\varrho \bb{u}\bb{u} + \left(p_\perp + n T_{\rm e} +\frac{B^2}{8\pi} \right)\msb{I} \right. \nonumber\\
    &\hspace{0.5in} \mbox{} - \left.\left(\frac{B^2}{4\pi} + \Delta p\right)\hat{\bb{b}}\hat{\bb{b}} \right] + \bb{F}.\label{eq:app_euler}
\end{align}
Similarly, the evolution of magnetic energy follows from Faraday's and Ohm's laws:
\begin{align}
    \frac{\partial \bb{B}}{\partial t} &= - c \grad\btimes\bb{E}  \label{eq:app_faraday}\\
    &= \grad \btimes \left[ \bb{u}\btimes\bb{B} - \frac{\left(\grad \btimes \bb{B}\right)\btimes \bb{B}}{4\pi n e/c}  \right. \nonumber \\
    &\hspace{0.95in} \left. \mbox{} + \eta_{\rm hyper} \grad^2 \left(\frac{\grad \btimes \bb{B}}{4\pi/c} \right) \right], \nonumber
\end{align}
where the last term represents dissipation due to hyper-resistivity. Equations for the evolution of kinetic and magnetic energies can be obtain by multiplying equations (\ref{eq:app_euler}) and (\ref{eq:app_faraday}) by $\bb{u}$ and $\bb{B}$ correspondingly. Thermal energy $\mathcal{E}_{\rm th}$ in the system increases due to viscous dissipation and compressive heating:
\begin{align}
    \frac{{\rm d} \mathcal{E}_{\rm th}}{{\rm d} t} &= -\int \msb{P}\,\bb{:}\,\grad \bb{u} \,{\rm d}^3 \bb{x}  \nonumber\\
    &= -\int \left[ p_\perp \grad\bcdot \bb{u} - (\Delta p\, \hat{\bb{b}}\hat{\bb{b}})\,\bb{:}\,\grad \bb{u}\right]\,{\rm d}^3\bb{x}. \label{eq:app_heating}
\end{align}

In what follows, it is useful to define the energy ``reservoirs'' corresponding to kinetic, magnetic and thermal energies, with each reservoir associated with a certain vector field $\bb{a}$, so that $\mathcal{E}_{\boldsymbol{a}} \equiv \bb{a}^2/2$. For the bulk kinetic energy, $\bb{a}^u = \sqrt{\varrho} \bb{u}$, so
\begin{align}
    \mathcal{E}_{\rm bulk} &= \int \frac{\left(\sqrt{\varrho}\bb{u}\right)^2}{2}\,{\rm d}^3 \bb{x}  \nonumber\\ \label{eq:app_kin_energy}
    &=\frac{1}{(2\pi)^3} \int \frac{1}{2}\left(\sqrt{\varrho}\bb{u}\right)_{\bs{k}} \bcdot \left(\sqrt{\varrho}\bb{u}\right)_{\bs{k}}^*\,{\rm d}^3\bb{k}.
\end{align}
The definition for magnetic energy is also straightforward: $\bb{a}^B \equiv\bb{B}/\sqrt{4\pi}$, so 
\begin{align}
    \mathcal{E}_{\rm mag}  &= \int \frac{\left(\bb{B}/\sqrt{4\pi}\right)^2}{2}\,{\rm d}^3 \bb{x}  \nonumber\\
    &= \frac{1}{(2\pi)^3} \int \frac{1}{8\pi}\left(\bb{B}\right)_{\bs{k}} \bcdot \left(\bb{B}\right)_{\bs{k}}^*\,{\rm d}^3\bb{k}.
\end{align}
The definition of the energy reservoir associated with anisotropic thermal energy is less straightforward. In this article, we use $\bb{a}^{\Delta p} \equiv \sqrt{|\Delta p|} \hat{\bb{b}}$. This definition is motivated by two facts. First, in the KRMHD limit, $(\delta \bb{a}^{\Delta p})^2/2$ matches up to a sign with the corresponding term in the plasma free energy \citep{Kunz2015}:
\begin{equation}
    W_{\rm KRMHD}^{\Delta} = \int \beta_{\parallel \rm i} \frac{\Delta_{\rm i}}{2} \frac{\delta B_\perp^2}{B_0^2} \,{\rm d}^3 \bb{x}.
\end{equation}
Secondly, the thermal energy of the plasma can be written as
\begin{align}
    &\mathcal{E}_{\rm th} = \int \frac{n(2T_\perp + T_\parallel)}{2} \,{\rm d}^3 \bb{x} = \int\left( \frac{3}{2} p_\perp - \frac{1}{2}\Delta p\right) \,{\rm d}^3 \bb{x}  \nonumber\\
    & = \mathcal{E}_{\rm th}^{\rm iso} - \int \frac{\vartheta_{\Delta p}}{2}\left(\sqrt{|\Delta p|} \hat{\bb{b}} \right)^2\,{\rm d}^3 \bb{x}  \label{eq:app_th_energy}\\
    & = \mathcal{E}_{\rm th}^{\rm iso} - \frac{1}{(2\pi)^3}\int  \frac{\vartheta_{\Delta p}}{2}\left(\sqrt{|\Delta p|} \hat{\bb{b}} \right)_{\bs{k}} \bcdot \left(\sqrt{|\Delta p|} \hat{\bb{b}} \right)_{\bs{k}}^* \,{\rm d}^3 \bb{k},\nonumber
\end{align}
where $\vartheta_{\Delta p} \equiv {\rm sign}(\Delta p)$ and $\mathcal{E}_{\rm th}^{\rm iso}$ is the ``isotropic thermal energy'' associated with $p_\perp$. The proposed definition of the anisotropic thermal energy allows us to consider the ``advection-like'' terms in (\ref{eq:app_heating}) and (\ref{eq:app_kin_energy}) associated with $\Delta p$ separately from the ``compression-like'' terms proportional to divergences of various vector fields in the system and terms proportional to gradients of $p_\perp$ and $B^2/2$.

\begin{figure*}
    \centering
    \includegraphics[width=\textwidth]{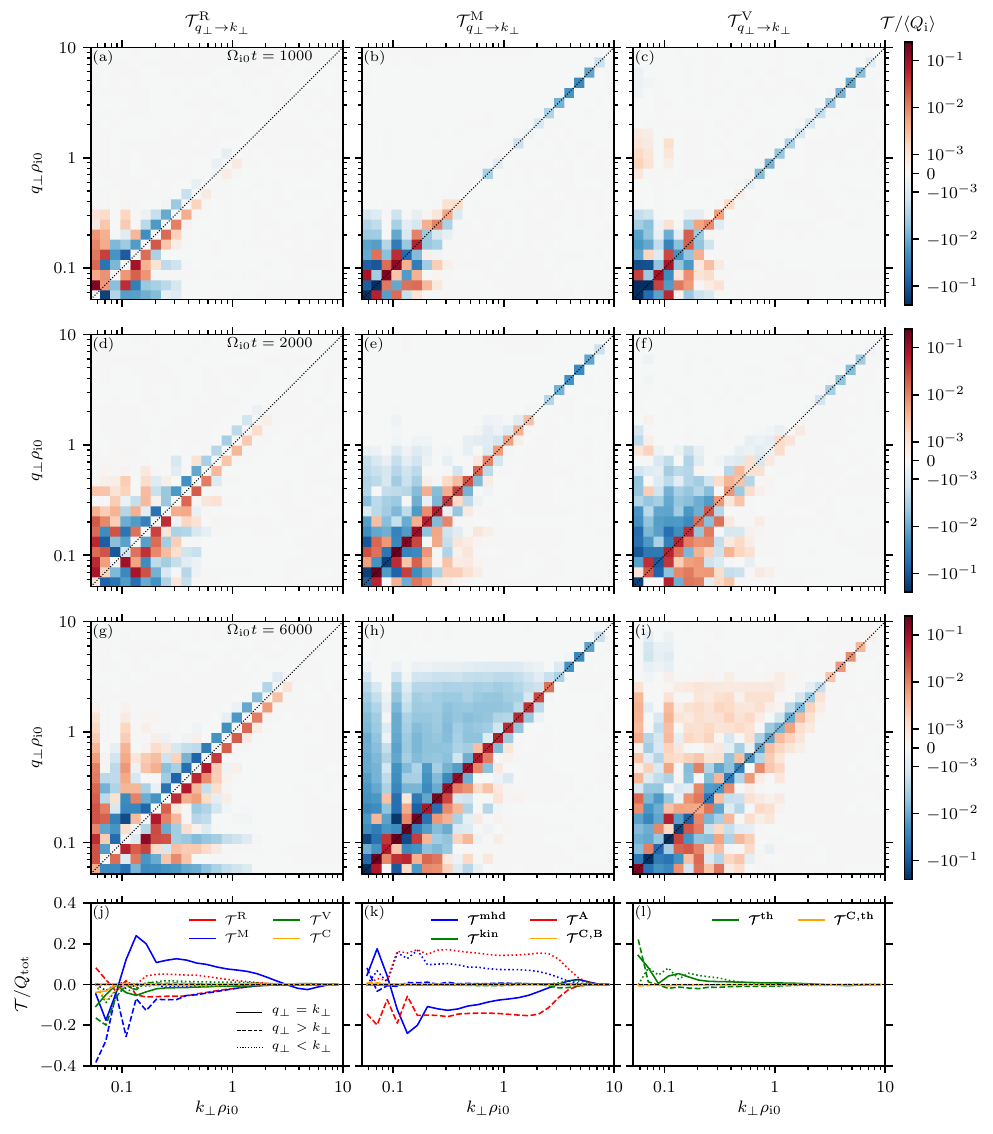}
    \caption{Non-local energy transfer functions averaged over the quasi-steady state due to the Reynolds stress (left), Maxwell stress (center) and anisotropic viscous stress (right) in $\beta_{\rm i0} = 16$ simulation. Most of the transfer in the early stages of the simulation is local, but there is considerable non-local transfer mediated by the Maxwell stress in the quasi-steady state. Panels (j)--(l) show the remaining transfer terms in the momentum equation as well as terms in the induction equation and the equation for thermal energy. Note that terms associated with compressions and gradients of isotropic pressures are smaller than ``advection-like'' terms.}
    \label{fig:nonlocal}
\end{figure*}

The rates of exchange of energy between different reservoirs follow from equations (\ref{eq:app_euler})--(\ref{eq:app_heating}). Let us consider the individual terms in equation (\ref{eq:app_euler}) separately. As we show later (Figure~\ref{fig:nonlocal}), the most dominant terms are related to the Reynolds, Maxwell and anisotropic viscous stresses. The rate of change of kinetic energy due to the Reynolds stress is
\begin{align}
    \frac{{\rm d} \mathcal{E}_{\rm bulk}^{\rm R}}{{\rm d} t} &= -\int \frac{(\sqrt{\varrho} \bb{u})}{\sqrt{\varrho}} \bcdot \left\{\grad\bcdot \left[\left(\sqrt{\varrho} \bb{u}\right) \left(\sqrt{\varrho} \bb{u}\right) \right]\right\} \,{\rm d}^3\bb{x} \nonumber\\
    = -&\int \frac{\bb{a}^u}{\sqrt{\varrho}} \bcdot \left[\grad \bcdot\left(\bb{a}^u \bb{a}^u \right)\right] \,{\rm d}^3\bb{x}  \\
    = -&\int \left\{\bb{a}^u \bcdot \left[\frac{\bb{a}^u}{\sqrt{\varrho}}\bcdot\grad\bb{a}^u\right] + \frac{(\bb{a}^u)^2}{\sqrt{\varrho}} \grad\bcdot\bb{a}^u \right\}\,{\rm d}^3\bb{x}. \nonumber
\end{align}
The second term in the integrand of the final expression is related to compressive motions (${\propto}\grad\bcdot \bb{a}^u$). As we show later (Figure~\ref{fig:nonlocal}), in high-$\beta$ turbulence, this term is smaller than the first,``advection-like'' term. To explore the locality of this cascade, we define $k$-shell averaged fields
\begin{equation}
    \langle\bb{a}\rangle_K \equiv \frac{1}{(2\pi)^3} \int\limits_{\bs{k} \in K} \bb{a}_{\bs{k}} \,{\rm e}^{-{\rm i} \bs{k}\bcdot\bs{x}}\,{\rm d}^3\bb{k}.
\end{equation}
We choose cylindrical $k_\perp$-shells with width ${\rm d} k_\perp \propto k_\perp$, so that the shells have equal width in $\log k_\perp$. With such a choice of shells, $\langle\bb{a}\rangle_K$ is proportional to the fluctuation amplitude $\delta \bb{a}_{k_\perp}$ (i.e., $\delta \bb{a}^2_{k_\perp} = \int_{k_\perp} |\bb{a}_{\bs{k}}|^2 {\rm d}^3 \bb{k}$). As a result, an energy transfer rate of the form $\langle\bb{u}\rangle_K \bcdot \left[\langle\bb{u}\rangle_K \bcdot \grad \right] \langle\bb{u}\rangle_K \sim k_\perp \delta \bb{u}_{k_\perp}^3 \sim \varepsilon$ approximates the cascade rate. For a conservative cascade, this energy transfer rate is the same for each $k_\perp$-shell.

For non-overlapping shells, the shell-averaged vectors $\bb{a}_K$ satisfy by construction
\begin{equation}
    \bb{a} = \sum\limits_K \langle\bb{a}\rangle_K,
\end{equation}
and, therefore,
\begin{align}
    &\frac{{\rm d} \mathcal{E}_{\rm bulk}^{\rm R}}{{\rm d} t} = \sum\limits_K\int \frac{{\rm d} \langle\bb{a}^u\rangle_K^2}{{\rm d} t}\,{\rm d}^3\bb{x}  \\
    &\approx -\sum\limits_K\int \langle\bb{a}^u\rangle_K \bcdot \left(\bb{u}\bcdot \grad \bb{a}^u\right) \,{\rm d}^3\bb{x} \equiv \sum\limits_{KPQ} \int T_{KPQ}^{\rm R} \,{\rm d}^3\bb{x},\nonumber
\end{align}
where in the last step we have defined 3-shell correlators (see, e.g. \citep{Kraichnan1967} for details):
\begin{equation}
    T^{\rm R}_{KPQ} \equiv -\langle\bb{a}^u\rangle_K \bcdot \left(\langle\bb{u}\rangle_P\bcdot \grad\right) \langle\bb{a}^u\rangle_Q.
\end{equation}
These correlators describe the transfer of energy associated with field $\bb{a}^u$ from shell $Q$ to shell $K$, mediated by field $\langle\bb{u}\rangle_P$, i.e.,
\begin{equation}
    \int \left.\frac{{\rm d} \langle\bb{a}^u\rangle_K^2}{{\rm d}t}\right|_{PQ} \,{\rm d}^3\bb{x} = \int T^{\rm R}_{KPQ}\,{\rm d}^3\bb{x}.
\end{equation}
The energy cannot go from shell $P$ to shells $K$ and $Q$ or vice versa. This can be easily seen from the sum of the energies in the shells $K$ and $Q$ during their mutual interaction mediated by the shell $P$:
\begin{align}
    &\int \frac{{\rm d} \langle\bb{a}^u\rangle_K^2}{{\rm d} t}\biggr|_{PQ} + \frac{{\rm d} \langle\bb{a}^u\rangle_Q^2}{{\rm d} t}\biggr|_{PK}\,{\rm d}^3\bb{x}  \\
    &= -\int \left[\langle\bb{a}^u\rangle_K \bcdot \left(\langle\bb{u}\rangle_P\bcdot \grad \langle\bb{a}^u\rangle_Q\right) \right.\nonumber\\
    &\hspace{0.5in}\left. \mbox{} +\langle\bb{a}^u\rangle_Q \bcdot \left(\langle\bb{u}\rangle_P\bcdot \grad \langle\bb{a}^u\rangle_K  \right)\right]\,{\rm d}^3\bb{x}  \nonumber \\
    &= -\int \langle\bb{u}\rangle_P\bcdot \grad \langle\bb{a}^u\rangle_K \bcdot \langle\bb{a}^u\rangle_Q\,{\rm d}^3\bb{x}  \nonumber\\
    &= \int\langle\bb{a}^u\rangle_K \bcdot \langle\bb{a}^u\rangle_Q\grad\bcdot \langle\bb{u}\rangle_P\,{\rm d}^3\bb{x}, \nonumber
\end{align}
so that the total energy during the interaction can only come from the compressive motions of the mediator shell, which are small in high-$\beta$ turbulence. As we are not interested in specific mediator shells, we can sum over $P$ and introduce a {\it shell-to-shell} energy transfer function, i.e., the energy-transfer function due to the Reynolds stress \citep{Grete2017}:
\begin{equation}
    \mathcal{T}^{\rm R}_{q_\perp \rightarrow k_\perp} \equiv -\int \langle \bb{a}^u\rangle_{k_\perp} \bcdot \left(\bb{u}\bcdot \grad \langle \bb{a}^u \rangle_{q_\perp}\right) \,{\rm d}^3 \bb{x}.\label{eq:app_reyn}
\end{equation}

Similar transfer functions could be defined for any field $\bb{a}$ and any mediator function $\bb{f}$:
\begin{equation}
    \mathcal{T}^{a_1fa_2}_{q_\perp \rightarrow k_\perp} \equiv \int \langle \bb{a}_1\rangle_{k_\perp} \bcdot \left( \bb{f}\bcdot \grad \langle \bb{a}_2 \rangle_{q_\perp}\right)\,{\rm d}^3 \bb{x}.
\end{equation}
Of particular importance to us are the transfer functions rates due to the Reynolds stress (\ref{eq:app_reyn}), the Maxwell stress,
\begin{equation}
    \mathcal{T}^{\rm M}_{q_\perp \rightarrow k_\perp} \equiv \int \langle \bb{a}^u\rangle_{k_\perp} \bcdot\left( \frac{\bb{B}}{\sqrt{4\pi \varrho}}\bcdot \grad \langle \bb{a}^B \rangle_{q_\perp}\right)\,{\rm d}^3 \bb{x},\label{eq:app_maxw}
\end{equation}
and the anisotropic viscous stress,
\begin{equation}
    \mathcal{T}^{\rm V}_{q_\perp \rightarrow k_\perp} \equiv \int \langle \bb{a}^u\rangle_{k_\perp} \bcdot \left(\vartheta_{\Delta p} \sqrt{\frac{|\Delta p|}{\varrho}} \hat{\bb{b}}\bcdot \grad \langle \bb{a}^{\Delta p} \rangle_{q_\perp}\right)\,{\rm d}^3 \bb{x}.\label{eq:app_delp}
\end{equation}
The mediator field for the Maxwell stress has a form of local Alfv\'en speed, and the mediator for the viscous stress has the form of the sound speed directed along the local magnetic field, but computed using the pressure anisotropy $\Delta p$. Both transfer functions have corresponding terms (of opposite signs) in the equations for magnetic energy and the anisotropic thermal energy. Namely,
\begin{align}
    \left.\frac{{\rm d} \mathcal{E}_{\rm mag}}{{\rm d} t}\right|_{\rm MHD} &= \int\frac{1}{4\pi} \bb{B} \bcdot\left[\grad\btimes(\bb{u}\btimes\bb{B})\right] \,{\rm d}^3\bb{x}  \\
    & = \int \left[\bb{a}^B\bcdot\left(\bb{a}^B\bcdot\grad \bb{u}\right) - \bb{u}\bcdot\grad(\bb{a}^B)^2/2  \right.\nonumber \\
    &\hspace{0.5in}\left. \mbox{} - (\bb{a}^B)^2 \grad\bcdot\bb{u}\right]\,{\rm d}^3 \bb{x}  \nonumber\\
    & = \sum\limits_{q_\perp k_\perp}\mathcal{T}^{\rm mhd}_{q_\perp\rightarrow k_\perp} + (\text{compressive terms}), \nonumber
\end{align}
where we only used the ``MHD'' electric field $\bb{u}\btimes\bb{B}/c$ and 
\begin{equation}
    \mathcal{T}^{\rm mhd}_{q_\perp \rightarrow k_\perp} \equiv \int \langle \bb{a}^B\rangle_{k_\perp} \bcdot \left( \frac{\bb{B}}{\sqrt{4\pi \varrho}}\bcdot \grad \langle \bb{a}^u \rangle_{q_\perp}\right)\,{\rm d}^3 \bb{x} ,
\end{equation}
and
\begin{align}
    \frac{{\rm d} \mathcal{E}^{\rm V}_{\rm th}}{{\rm d} t} &= \int \Delta p \hat{\bb{b}}\hat{\bb{b}}\,\bb{:}\,\grad \bb{u} \,{\rm d}^3\bb{x}  \\
    &=\int \sqrt{\varrho}\bb{a}^{\Delta p} \bcdot \left(\bb{a}^{\Delta p}\bcdot \grad \bb{u}\right)  \nonumber \\
    & = \sum\limits_{q_\perp k_\perp}\mathcal{T}^{\rm th}_{q_\perp\rightarrow k_\perp} + (\text{compressive terms}),\nonumber
\end{align}
where 
\begin{equation}
    \mathcal{T}^{\rm th}_{q_\perp \rightarrow k_\perp} \equiv \int \langle \bb{a}^{\Delta p}\rangle_{k_\perp} \bcdot \left( \vartheta_{\Delta p}\sqrt{\frac{|\Delta p|}{\varrho}} \hat{\bb{b}}\bcdot \grad \langle \bb{a}^u \rangle_{q_\perp} \right) \,{\rm d}^3\bb{x}.
\end{equation}

To summarize, we write the time derivative of the bulk kinetic energy as a sum of five terms:
\begin{align}
    \frac{{\rm d} \mathcal{E}_{\rm bulk}}{{\rm d} t} &= \sum\limits_{q_\perp k_\perp} \left[ \mathcal{T}^{\rm R}_{q_\perp\rightarrow k_\perp} + \mathcal{T}^{\rm M}_{q_\perp\rightarrow k_\perp} + \mathcal{T}^{\rm V}_{q_\perp\rightarrow k_\perp}\right. \\
    &\hspace{0.5in} \left.\mbox{} + \mathcal{T}^{\rm C}_{q_\perp\rightarrow k_\perp} + \mathcal{T}^{\rm F}_{q_\perp\rightarrow k_\perp}\right],\nonumber
\end{align}
where $\mathcal{T}^{\rm F}_{q_\perp\rightarrow k_\perp}$ represents energy injection due to large-scale external forcing, and $\mathcal{T}^{\rm C}_{q_\perp\rightarrow k_\perp}$ is the sum of all terms neglected in other transfer functions, which are related to compressive motions and various pressure forces:
\begin{align}
    \mathcal{T}^{\rm C}_{q_\perp\rightarrow k_\perp} \equiv \int\biggl\{ &- \langle\bb{a}^u\rangle_{k_\perp} \bcdot \langle\bb{a}^u\rangle_{q_\perp} \frac{\grad\bcdot \bb{u}}{2}  \\
    & \mbox{} - \langle\bb{a}^u\rangle_{k_\perp} \bcdot \frac{1}{\sqrt{\varrho}} \grad \langle p_\perp + n T_{\rm e}\rangle_{q_\perp}  \nonumber\\
    & \mbox{}  - \langle\bb{a}^u\rangle_{k_\perp} \bcdot \left[ \frac{1}{2\sqrt{4\pi\varrho}} \grad \left(\bb{B} \bcdot \langle\bb{a}^B\rangle_{q_\perp} \right)\right] \nonumber\\
    & \mbox{}  + \langle\bb{a}^u \rangle_{k_\perp} \bcdot \langle\bb{a}^{\Delta p}  \rangle_{q_\perp} \vartheta_{\Delta p}\frac{\grad \bcdot \sqrt{|\Delta p|} \hat{\bb{b}}}{\sqrt{\varrho}}\biggr\}\,{\rm d}^3 \bb{x}.\nonumber
\end{align}

The magnetic energy evolves according to a similar sum of energy-transfer functions:
\begin{align}
    \frac{{\rm d} \mathcal{E}_{\rm mag}}{{\rm d} t} &= \sum\limits_{q_\perp k_\perp} \left[ \mathcal{T}^{\rm mhd}_{q_\perp\rightarrow k_\perp} + \mathcal{T}^{\rm kin}_{q_\perp\rightarrow k_\perp}  + \mathcal{T}^{\rm A}_{q_\perp\rightarrow k_\perp}\right. \\
    & \hspace{0.5in}\left.\mbox{} + \mathcal{T}^{\rm C,B}_{q_\perp\rightarrow k_\perp}+ \mathcal{T}^{\rm diss}_{q_\perp\rightarrow k_\perp}\right],\nonumber
\end{align}
where
\begin{align}
    \mathcal{T}^{\rm mhd}_{q_\perp\rightarrow k_\perp} \equiv \int & \langle\bb{a}^B\rangle_{k_\perp} \bcdot \left( \frac{\bb{B}}{\sqrt{4\pi\varrho}} \bcdot \grad \langle\bb{a}^u\rangle_{q_\perp}\right) \,{\rm d}^3 \bb{x}, \\
    \mathcal{T}^{\rm kin}_{q_\perp\rightarrow k_\perp} \equiv\int & - \langle\bb{a}^B\rangle_{k_\perp} \bcdot \left( \frac{\bb{B}}{\sqrt{4\pi\varrho}} \bcdot\grad \left\langle\frac{\sqrt{\varrho}\bb{J}}{n e}\right\rangle_{q_\perp} \hspace{-0.02in}\right)\,{\rm d}^3\bb{x}, \\
    \mathcal{T}^{\rm A}_{q_\perp\rightarrow k_\perp} \equiv \int &  -\langle\bb{a}^B\rangle_{k_\perp} \bcdot \left[ \left( \bb{u} - \frac{\bb{J}}{n e} \right) \bcdot \grad \langle\bb{a}^B\rangle_{q_\perp} \right] {\rm d}^3 \bb{x},\\
    \mathcal{T}^{\rm C,B}_{q_\perp\rightarrow k_\perp} \equiv \int & \biggl\{- \langle\bb{a}^B\rangle_{k_\perp} \bcdot \langle\bb{a}^B\rangle_{q_\perp} \grad\bcdot\bb{u}  \\
    &\mbox{} + \langle\bb{a}^B\rangle_{k_\perp} \bcdot \langle\bb{a}^B\rangle_{q_\perp} \grad\bcdot\frac{\bb{J}}{n e} \nonumber\\
     \mbox{} - \frac{1}{2}  \langle\bb{a}^B\rangle_{k_\perp} \bcdot  & \left(  \langle \bb{a}^u\rangle_{q_\perp}  - \left\langle\frac{\sqrt{\varrho}\bb{J}}{n e}\right\rangle_{q_\perp}\right) \frac{\bb{B}}{\sqrt{4\pi\varrho}}\bcdot\frac{\grad \varrho}{\varrho}\biggr\} \,{\rm d}^3 \bb{x}, \nonumber \\
    \mathcal{T}^{\rm diss}_{q_\perp\rightarrow k_\perp} \equiv \int& \frac{\langle\bb{a}^B\rangle_{k_\perp}}{\sqrt{4\pi}} \bcdot\left[ \grad\btimes\left(\grad^2 \left\langle\frac{\eta_{\rm hyper}\bb{J}}{n e}\right\rangle_{q_\perp} \right) \right]{\rm d}^3 \bb{x}.
\end{align}

Finally, the thermal energy evolves according to 
\begin{equation}
    \frac{{\rm d} \mathcal{E}_{\rm th}}{{\rm d} t} = \sum\limits_{q_\perp k_\perp} \left[ \mathcal{T}^{\rm th}_{q_\perp\rightarrow k_\perp} + \mathcal{T}^{\rm C,th}_{q_\perp\rightarrow k_\perp} \right],
\end{equation}
where
\begin{align}
    \mathcal{T}^{\rm th}_{q_\perp\rightarrow k_\perp} &\equiv \int\langle\bb{a}^{\Delta p} \rangle_{k_\perp} \bcdot \left( \vartheta_{\Delta p}\sqrt{\frac{|\Delta p|}{\varrho}}\hat{\bb{b}}\bcdot \grad \langle\bb{a}^{u} \rangle_{q_\perp}\right) \,{\rm d}^3 \bb{x},\\
    \mathcal{T}^{\rm C,th}_{q_\perp\rightarrow k_\perp} &\equiv \int\left\{ - \frac{\grad\bcdot\langle\bb{a}^u\rangle_{k_\perp}}{\sqrt{\varrho}} \langle p_\perp\rangle_{q_\perp} \right.\\
    &\mbox{} - \frac{\langle\bb{a}^u\rangle_{k_\perp}\bcdot\langle\bb{a}^u\rangle_{q_\perp}}{2} \grad\bcdot \bb{u} \nonumber\\
    &\mbox{} \left. -\frac{1}{2} \langle\bb{a}^{\Delta p}\rangle_{k_\perp} \bcdot \langle\bb{a}^u\rangle_{q_\perp} \vartheta_{\Delta p}\sqrt{\frac{|\Delta p|}{\varrho}}\hat{\bb{b}}\bcdot \frac{\grad \varrho}{\varrho}\right\}\,{\rm d}^3 \bb{x}. \nonumber
\end{align}

The transfer functions $\mathcal{T}^{\rm R}_{q_\perp\rightarrow k_\perp}$,$\mathcal{T}^{\rm M}_{q_\perp\rightarrow k_\perp}$ and $\mathcal{T}^{\rm V}_{q_\perp\rightarrow k_\perp}$ for our $\beta_{\rm i0} = 16$ simulation are shown in Figure \ref{fig:nonlocal}. The energy transfer due to Reynolds stress is consistent with a local cascade (see, e.g. \citep{Grete2017}): there is a positive energy flux coming from large scales to small scales ($q_\perp < k_\perp$) and a similar negative flux from small scales to large scales ($q_\perp > k_\perp$). The energy transfers due to Maxwell and viscous stresses are mostly local: there is a positive energy-transfer rate at $q_\perp = k_\perp$ coming from the magnetic tension, and a negative transfer mediated by pressure anisotropy at the same $q_\perp$. However, unlike in a local MHD cascade, there is a considerable energy transfer at $q_\perp > k_\perp$ mediated by the Maxwell stress. This corresponds to the transfer of large-scale kinetic energy to small-scale magnetic energy, as discussed in \S\ref{sec:energy_transfer}.

\begin{figure*}
    \centering
    \includegraphics[width=\textwidth]{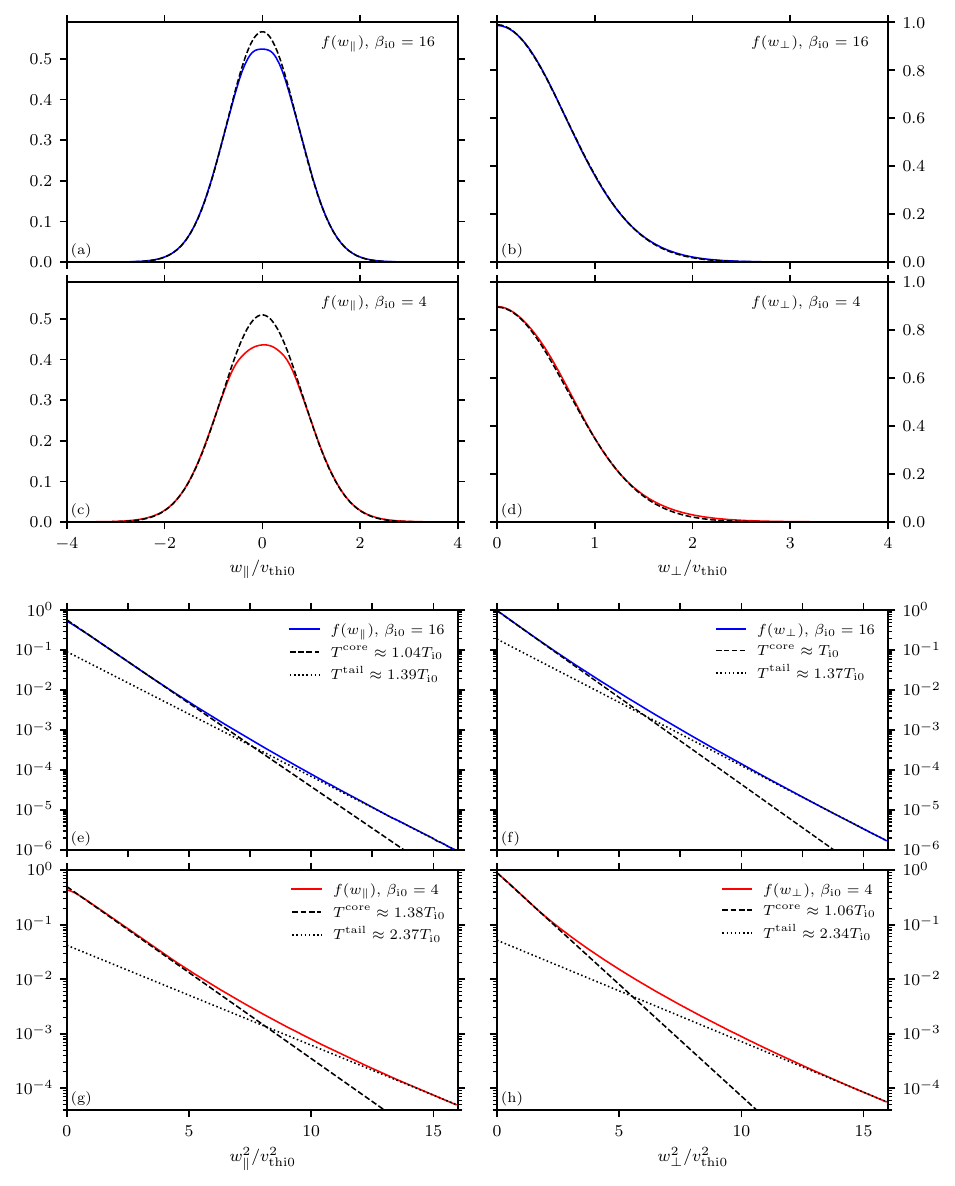}
    \caption{Parallel (a,c) and perpendicular (b,d) ion distribution function at the end of the $\beta_{\rm i0} = 16$ (a,b) and $\beta_{\rm i0} = 4$ (c,d) simulations. Parallel distribution functions (a,c) have a flat core, which we associate with the Landau damping and the firehose instability~\citep{Bott2021}. Dashed lines show best-fit Maxwellians to $f(w_\parallel)$ at $w_\parallel \approx 1.6 v_{\rm thi0}$ (in particular, from $1.59$ to $1.61 v_{\rm thi0}$ to exclude the flattened part of the core) and to $f(w_\perp)$ at $w_\perp < v_{\rm thi0}$. (e--h) The same distribution functions, plotted on logarithmic scale. The core and the tail of the distributions may be approximated with gaussians (straight lines in these coordinates). It is apparent that the wings of the distribution functions are much more isotropic than the cores ($T^{\rm tail}_\parallel \sim T^{\rm tail}_\perp$, while $T^{\rm core}_\parallel > T^{\rm core}_\perp$).}
    \label{fig:dist}
\end{figure*}

Finally, we note that the largest, ``advection-like'' energy-transfer terms correspond to turbulent dynamics and micro-fluctuations produced by instabilities. These terms, however, do not describe thermalization of anisotropic thermal energy due to the effective collisionality. Studying this thermalization requires considering the full equation for pressure anisotropy, which involves higher-order moments of the distribution functions, such as the heat fluxes. Such an analysis goes beyond the scope of this paper.

\section{Ion distribution function}\label{sec:appendix_dist}

Figure~\ref{fig:dist} shows the ion parallel and perpendicular distribution functions at the end of the  $\beta_{\rm i0} = 4$ and $\beta_{\rm i0} = 16$ runs, defined as $f(w_\parallel) \equiv \int {\rm d}w_\perp^2\, f(w_\parallel,w_\perp)$ and $f(w_\perp) \equiv \int {\rm d}w_\parallel\, f(w_\parallel,w_\perp)$. Both runs produce distribution functions close to a bi-Maxwellian with $T_\parallel > T_\perp$. The core of the parallel distribution function is flattened, which we attribute to Landau damping (mostly active in the early stages of the simulations) and to the non-linear phase of the firehose instability \footnote{A similar flattened-core feature was seen previously in a kinetic simulation of expanding Alfv\'{e}nic turbulence, which was also affected by the firehose instability but exhibited significantly less Landau damping~\citep{Bott2021} because of its lower value of $\beta_{\rm i0}$.}.

Although the final distribution functions are not isotropic, the tails of those distributions are, as we show in Figure~\ref{fig:dist}(e--h). This implies that pitch-angle scattering due to kinetic instabilities affects different parts of the distribution function differently, and causes the wings of the distribution function to be much more collisional, and therefore isotropic, than the cores. Similar features had been seen in the distribution functions of prior $\beta\sim1$ simulations~\citep{Arzamasskiy2019}, in which pitch-angle scattering was present without instabilities, and in simulations of magnetized turbulence in an expanding box~\citep{Bott2021}, in which anomalous scattering was associated with firehose fluctuations. Understanding such distribution functions requires careful examination of the particle-energy dependence of the effective collisionality, which falls beyond the scope of this paper.

\end{document}